\documentclass[a4paper,fleqn,usenatbib]{mn2e}

\usepackage[T1]{fontenc}
\usepackage{ae,aecompl}

\long\def\symbolfootnote[#1]#2{\begingroup%
\def\thefootnote{\fnsymbol{footnote}}\footnote[#1]{#2}\endgroup}

\usepackage{graphicx}	
\usepackage{amsmath}	
\usepackage{amssymb}	
\usepackage{xcolor}
\usepackage{pdflscape} 

\newcommand{\ha}{H$\alpha$~}
\newcommand{\hb}{H$\beta$~}
\newcommand{\hans}{H$\alpha$}
\newcommand{\hbns}{H$\beta$}

\newcommand\msun{{{{\rm M_\odot}}}}

\newcommand{\mgii}{Mg\,{\sc ii}~}
\newcommand{\mgiins}{Mg\,{\sc ii}}
\newcommand{\feii}{Fe\,{\sc ii}~}

\newcommand{\civ}{C\,{\sc iv}~}
\newcommand{\civns}{C\,{\sc iv}}
\newcommand{\ciii}{C\,{\sc iii}]~}

\newcommand{\heiins}{He\,{\sc ii}}
\newcommand{\feiins}{Fe\,{\sc ii}}

\newcommand{\oiv}{O\,{\sc iv}~}
\newcommand{\oivns}{O\,{\sc iv}}

\newcommand{\siivns}{Si\,{\sc iv}}
\newcommand{\oiii}{[O\,{\sc iii}]~}
\newcommand{\oiiins}{[O\,{\sc iii}]}
\newcommand{\oi}{O\,{\sc i}~}
\newcommand{\siii}{[Si\,{\sc ii}]~}
\newcommand{\siiii}{Si\,{\sc iii}]~}

\newcommand{\niins}{[N\,{\sc ii}]}

\newcommand{\siins}{[S\,{\sc ii}]}

\def \ll {$\tt\lambda\lambda$}
\def \l {$\tt \lambda$}

\def \kms {km\,s$^{-1}$}

\newcommand{\ebv}{\ensuremath{E(B-V)}}

\title[\civns-Emission Line Properties]{C\,{\sc iv} Emission Line Properties and Systematic Trends in Quasar Black Hole Mass Estimates}

\author[L. Coatman et al.]{
Liam Coatman,$^{1}$\thanks{E-mail: lc585@ast.cam.ac.uk}
Paul C. Hewett,$^{1}$
Manda Banerji$^{1}$ and
Gordon T. Richards$^{2}$
\\
$^{1}$Institute of Astronomy, Madingley Road, Cambridge CB3 0HA, UK\\
$^{2}$Department of Physics, Drexel University, 3141 Chestnut Street, Philadelphia PA 19104, USA
}

\date{Accepted XXX. Received YYY; in original form ZZZ}

\pubyear{2016}

\begin{document}
\label{firstpage}
\pagerange{\pageref{firstpage}--\pageref{lastpage}}
\maketitle

\begin{abstract} 

Black-hole masses are crucial to understanding the physics of the connection between quasars and their host galaxies and measuring cosmic black hole-growth. 
At high redshift, $z \gtrsim 2.1$, black hole masses are normally derived using the velocity-width of the \civns\ll1548,1550 broad emission line, based on the assumption that the observed velocity-widths arise from virial-induced motions.  
In many quasars, the \civns-emission line exhibits significant blue asymmetries (`blueshifts') with the line centroid displaced by up to thousands of \kms\, to the blue. 
These blueshifts almost certainly signal the presence of strong outflows, most likely originating in a disc wind.
We have obtained near-infrared spectra, including the \ha\l6565 emission line, for 19 luminous ($L_{\rm Bol} = 46.5-47.5$ erg~s$^{-1}$) Sloan Digital Sky Survey quasars, at redshifts $2 < z < 2.7$, with \civ emission lines spanning the full-range of blueshifts present in the population.  
A strong correlation between \civns-velocity width and blueshift is found and, at large blueshifts, $>$2000\,\kms, the velocity-widths appear to be dominated by non-virial motions. 
Black-hole masses, based on the full width at half maximum of the \civns-emission line, can be overestimated by a factor of five at large blueshifts. 
A larger sample of quasar spectra with both \civ and \hbns, or \hans, emission lines will allow quantitative corrections to \civns-based black-hole masses as a function of blueshift to be derived. 
We find that quasars with large \civ blueshifts possess high Eddington luminosity ratios and that the fraction of high-blueshift quasars in a flux-limited sample is enhanced by a factor of approximately four relative to a sample limited by black hole mass.    

\end{abstract}




\section{Introduction}\label{sec:introduction}

Super-massive black holes (BHs) are found at the centres of most nearby massive galaxies and the BH mass and mass of the host galaxy spheroid are strongly correlated \citep{ferrarese00,gebhardt00,kormendy13}. 
Although any underlying causal mechanism(s) responsible for the correlation is yet to be conclusively identified, there is considerable observational and theoretical support for models that involve BH-fuelling, outflows and a `feedback' relationship \citep[e.g.][]{king15}.  
The number density of quasars, which evolves strongly with redshift, peaks at redshifts $2 \lesssim z \lesssim 3$ \citep[e.g.][]{brandt05,richards06b} and the most massive (M$_{\rm BH} \gtrsim 10^9\msun$) present-day BHs experienced much of their growth during this epoch.  
The star formation rate, which closely follows the cosmological evolution of the quasar luminosity function, also peaks during this epoch \citep[e.g.][]{boyle98}. 
Quantifying the growth-rate of massive BHs at $2 \lesssim z \lesssim 3$ would therefore help significantly in understanding the role quasars play in galaxy evolution.

Reliable estimates of BH masses are a prerequisite for investigating the relationship between BHs and their host galaxies.  
If the line-emitting clouds in the broad line region (BLR) are assumed to be virialized and moving in a potential dominated by the central BH, then the BH mass is simply a product of the BLR size and the square of the virial velocity.
The reverberation-mapping technique uses the time lag between variations in the continuum emission and correlated variations in the broad line emission to measure the typical size of the BLR \citep{peterson93,peterson14}. 
The full width at half maximum (FWHM) or dispersion ($\sigma$; derived from the second moment) velocity of the prominent broad emission line of \hb (4862.7\AA)\footnote{Vacuum wavelengths are employed throughout the paper.} is used as an indicator of the virial velocity, with extensions to other low-ionization emission lines such as \ha (6564.6\AA) and \mgiins\ll2796.4,2803.5 \citep[e.g.][]{vestergaard02,mclure02,wu04,kollmeier06,onken08,wang09,rafiee11}.
Extensive reverberation mapping campaigns have provided accurate BH masses for $\sim$50 active galactic nuclei (AGN) at relatively low redshifts and of modest luminosity \citep[e.g.][]{kaspi00,kaspi07,peterson04,bentz09,denney10}. 

Reverberation mapping campaigns have also revealed a tight relationship between the radius of the BLR and the quasar optical (or ultraviolet) luminosity \citep[the $R-L$ relation; e.g.][]{kaspi00,kaspi07}.
This relation provides a much less expensive method of measuring the BLR radius, and large-scale studies of AGN and quasar demographics have thus become possible through the calibration of single-epoch virial-mass estimators using the reverberation-derived BH masses \citep[e.g.][]{greene05,vestergaard06,vestergaard09,shen11,shen12,trakhtenbrot12}.
The uncertainties in reverberation mapped BH masses are estimated to be $\sim 0.4$ dex \citep[e.g.][]{peterson10}, and the uncertainties in virial masses are similar \citep[e.g.][]{vestergaard06}.
Since the structure and geometry of the BLR is unknown, a virial coefficient $f$ is introduced to transform the observed line-of-sight velocity inferred from the line width in to a virial velocity.
This simplification accounts for a significant part of the uncertainty in virial BH masses \citep[in addition to, for example, describing the BLR with a single radius $R$ and scatter in the $R-L$ relation;][]{shen13}. 
Furthermore, if the BLR is anisotropic \citep[for example, in a flattened disk; e.g.][]{jarvis06} then the line width will be orientation-dependent \citep[e.g.][]{runnoe13b,shen14,brotherton15}. 

At redshifts of $z\gtrsim 2.0$ the low-ionization hydrogen and \mgii emission lines are no longer present in the optical spectra of quasars and it is necessary to employ an emission line in the rest-frame ultraviolet.  
The strong \civns\ll1548.2,1550.8 emission doublet is visible in the optical spectra of quasars to redshifts of $z\sim5$ and \civns-derived BH masses have become the standard \citep[e.g.][]{vestergaard06,park13} for both individual quasars and in studies of quasar population demographics.

The luminosities of quasars at redshifts $z\gtrsim 2$ are much greater than the majority of AGN at lower redshifts for which reverberation mapping results are available.  
Therefore, the reliability of the existing calibration involving \civ FWHM velocity measurements and ultraviolet luminosity is not established definitively when extrapolating to high-redshifts and luminosities. 
While some authors have found good agreement between BH mass-estimates based on \civ and \hb \citep[e.g.][]{vestergaard06, assef11, tilton13}, others have questioned the consistency \citep[e.g.][]{baskin05,trakhtenbrot12,shen12}.  

\citet{denney12} presented evidence that the interpretation of the FWHM velocity of the \civns-emission being due primarily to virial motions within the quasar BLR requires care.  
Specifically, both a low-velocity core component and a blue excess to the \civns-emission, both of which do not reverberate, can be present and \citet{denney12} proposes that a contribution from an accretion disc wind or from a more distant narrow emission line region is important.

Certainly, in contrast to the hydrogen Balmer lines and \mgiins, the \civ emission line in quasar spectra exhibits a broad range of line shapes, including significant asymmetry, with shifts of the line-centroid to the blue (`blueshifts') of up to several thousand \kms\, \citep{richards02,baskin05,sulentic06}.  
\citet{shen12} found, using a sample of 60 luminous quasars, that the scatter between the FWHM of \civ and \hb was correlated with the blueshift of \civ relative to \hbns. 
\citet{shen08} found a similar result by comparing \civ with \mgii for quasars from SDSS DR5.  
The blueshifting of \civ is usually interpreted as evidence for strong outflows \citep[e.g.][]{sulentic07, richards11} which, most likely, result from the presence of a radiation line-driven accretion-disc wind \citep[e.g.][]{konigl94, murray95, proga00, everett05, gallagher15}.  
In this picture, the non-virial wind component makes a significant contribution to the observed \civns-emission FWHM in quasars with large \civ blueshifts (`wind-dominated quasars') and hence increases the inferred BH masses.
A primary goal of this paper is to present the full range of \civns-emission line blueshifts present among high-luminosity quasars at redshifts $z\sim2.5$ and investigate potential systematic trends in the derived \civns-based BH masses as a function of blueshift.

Changes in the \civ blueshift and equivalent width are correlated with changes in the velocity widths and strengths of other optical and ultra-violet emission lines.
In the spectra of lower-redshift AGN, the FWHM of the broad \hb emission line and the relative strengths of optical \feii and \hb have been identified as the features responsible for the largest variance in the population. 
These parameters form part of `Eigenvector 1' (EV1), the first eigenvector in a principal component analysis which originated from the work of \citet{boroson92}.   
The underlying driver behind EV1 is thought to be the Eddington ratio \citep[e.g.][]{sulentic00b,shen14}. 
\citet{sulentic00} proposed a two-population model to classify AGN by their EV1 properties. 
In this scheme AGN with FWHM(\hbns) < 4000 \kms\, and FWHM(\hbns) > 4000 \kms\, are classified as population A and B objects respectively, although there is a continuous distribution of parameter values across this divide. 
\citet{sulentic07} added a measure of the \civ asymmetry to EV1, and found a strong association between blue-asymmetry and their population A quasars.

\citet{denney12} found the level of contamination in single-epoch spectra from non-reverberating gas to be correlated with the shape (FWHM/$\sigma$) of the \civ profile. 
\citet{runnoe13} found the scatter between the \civ and \hb line widths to be correlated with the continuum-subtracted peak flux ratio of the ultraviolet emission-line blend of \siivns+\oiv (at 1400\,\AA) to that of \civns. 
Both authors used these correlations to propose empirical corrections to the \civ line width which can improve the consistency between \civ and \hbns-based virial BH mass estimates. 
In fact, the shape, peak flux relative to the 1400\,\AA\, blend, and blueshift of \civ all correlate with one another and with other parameters in EV1.
Therefore, EV1 provides a useful context for understanding systematic trends in \civ velocity widths, and hence virial BH masses. 

Currently, the number of reverberation mapped quasars is both small \citep[$\sim$50 quasars;][]{park13} and, as highlighted by \citet{richards11}, includes a restricted range of the \civ emission line shapes seen in the quasar population. 
In particular, the reverberation mapped objects generally possess high \civ equivalent widths and low \civns-blueshifts. 
Nevertheless, the derived scaling relations based on the reverberation-mapped sample are regularly applied to the quasar population with low \civ EWs and/or large \civns-blueshifts, where any non-virial outflow-related contribution to the dynamics is significant. 
Much more complete coverage of the \civns-emission properties within the population of luminous quasars will come from the new SDSS-IV reverberation mapping project \citep{shen15} but, for now, additional direct comparison of \civns-emission and low-ionization emission-line properties in the same quasars offers a way forward.

Near-infrared spectra, including the \ha emission line, for a sample of 19 quasars, at redshifts $2.0 < z < 2.7$, have been obtained to complement existing SDSS optical spectra covering the \civ emission line. 
The 19 quasars were chosen to include a broad range of \civ line blueshifts.
Our aim is to directly test the reliability of \civns-based BH mass estimates at high redshift for objects with a diverse range of \civns-line shapes.  
In particular, we will investigate potential systematic effects on the \civns-emission based BH masses for quasars with large, $\gtrsim$1200\kms, \civ blueshifts, using the properties of the \ha emission line to provide BH-mass estimates for the objects unbiased by non-virial contributions to the emission-line profile.
Examining higher redshifts, our work complements other studies which attempt to improve the reliability of BH mass estimates which use the \civ line \citep[e.g.][]{runnoe13,denney12}. 
However, the range of \civ blueshifts in our sample is significantly more extended, which will allow us to study systematic biases in \civns-based virial BH masses more directly, i.e. as a function of the \civ blueshift. 
Established relations to derive BH masses from emission-line properties are employed but an advantage of our approach is that \civ and \ha can be directly compared as a function of \civns-emission line shape.  
In Section~\ref{sec:selection} we describe how quasars were selected for rest-frame optical spectroscopy, before reviewing the \civ emission-line properties of high luminosity quasars in the SDSS surveys in Section~\ref{sec:blueshifts}. 
The new near-infrared spectroscopic observations are outlined in Section~\ref{sec:observations}. 
With spectroscopic coverage of the \civ and \ha emission lines available, the procedures used to quantify the line parameters are described in Section~\ref{sec:line_measurements}. 
The results of the spectral analysis in the context of BH masses and \civns-blueshift are presented in Section~\ref{sec:results}, with a discussion of the significance of the systematic trends identified included in Section~\ref{sec:discussion}.
The paper concludes with a summary of the main conclusions in Section~\ref{sec:conclusions}.
Throughout this paper we adopt a $\Lambda$CDM cosmology with $h_0=0.71$, $\Omega_M=0.27$, and $\Omega_\Lambda=0.73$. 
All wavelengths and equivalent width measurements are given in the quasar rest-frame, and all emission line wavelengths are given as measured in vacuum.

\section{Sample selection}
\label{sec:selection}

The parent sample for our investigation is the spectroscopic quasar catalogue of the Sloan Digital Sky Survey \citep[SDSS;][]{york00} Seventh Data Release \citep[DR7;][]{schneider10}. 
The SDSS DR7 catalogue contains moderate resolution $\sim3800-9180$\AA\, spectra for 105,783 quasars. 
\citet{shen11} have compiled a catalogue of properties for the SDSS DR7 quasars including, at $z > 1.5$, measurements of the broad \civns\ll1548.2,1550.8 emission line.
Our aim is to explore the relationship between the \ha and \civ emission-line properties over the full dynamic range in \civns-emission shapes, with particular emphasis on quasars possessing large \civns-blueshifts (see Section~\ref{sec:blueshifts}). 
The sample was restricted to objects with redshifts $2.14 < z <2.51$ (7,258 quasars), to ensure that the \hb and \ha emission lines fall within the $H$- and $K$-bands respectively, allowing us to observe both simultaneously with the appropriate grism configuration.
Given the limited number of quasars for which near-infrared spectra could be obtained, the quasar sample was further restricted to objects that are radio-quiet (5,980 quasars), show no evidence of broad absorption lines (BALs) in their spectra (5,299 quasars), and are free from significant dust extinction. 
We removed radio-loud objects from our sample using the same radio-loud classification as \citet{shen11}, and BAL quasars using the classifications of both \citet{shen11} and \citet{allen11}. 
The removal of quasars with significant dust extinction was achieved by identifying quasars with $i-K$ colours redder than a parametric spectral energy distributions (SED) model + SMC-like extinction curve with \ebv=0.05 \citep[see][]{maddox12}. 
The $K$-magnitude was taken from the UKIRT Infrared Deep Sky Survey \citep[UKIDSS;][]{lawrence07} Large Area Survey (ULAS). 
The requirement to be in the ULAS footprint and have reliable $K$ band photometry reduced our sample of possible targets to 1,683, and the \ebv\, cut left 1,204 in our sample. 
Finally, a flux-limit of $K<18.5$ (AB) was applied to ensure that spectra of sufficient signal-to-noise ratio (S/N) could be obtained (412 quasars). 
 
We were able to obtain new infra-red spectra for 19 quasars from this sample of 412 possible targets (Section \ref{sec:observations}). 
The quasars included in this sub-sample were selected to have \civns-emission shapes which span the full range observed in the population.
Reliably quantifying the distribution of \civns-emission shapes has been made possible thanks to redshift-determination algorithms \citep[][Allen \& Hewett 2016, in preparation]{hewett10} which are independent of the \civns-emission shape. 
Calculation of the \civ emission line parameters is described in detail in the next section. 

\section{\civ blueshifts in the quasar population}
\label{sec:blueshifts}

Recognition that the \civ emission line in quasars can exhibit significant asymmetric structure, with an excess of flux to the blue of the predicted rest-frame transition wavelength, extends back to \citet{gaskell82}. 
Significant progress in understanding the relationship between changes in \civns-emission shape and quasar properties has come about through studies in which near-infrared spectra of the hydrogen Balmer lines have been obtained. 
Such studies typically involve samples of modest size and the location of the Balmer lines provides a reliable estimate of the quasar systemic redshifts; recent examples include \citet{shen12} and \citet{marziani16}. 
In Section~\ref{sec:line_measurements} we adopt the same approach to estimate systemic redshifts for the quasar sample presented here with near-infrared spectra.  
However, improvements in the estimation of systemic redshifts from ultraviolet quasar spectra means that it is now possible to quantify the distribution of \civns-blueshifts in the observed quasar population as a whole. 

\subsection{Quasar systemic redshifts}
\label{sub:sysredshifts}

\begin{figure}
    \includegraphics[width=\columnwidth]{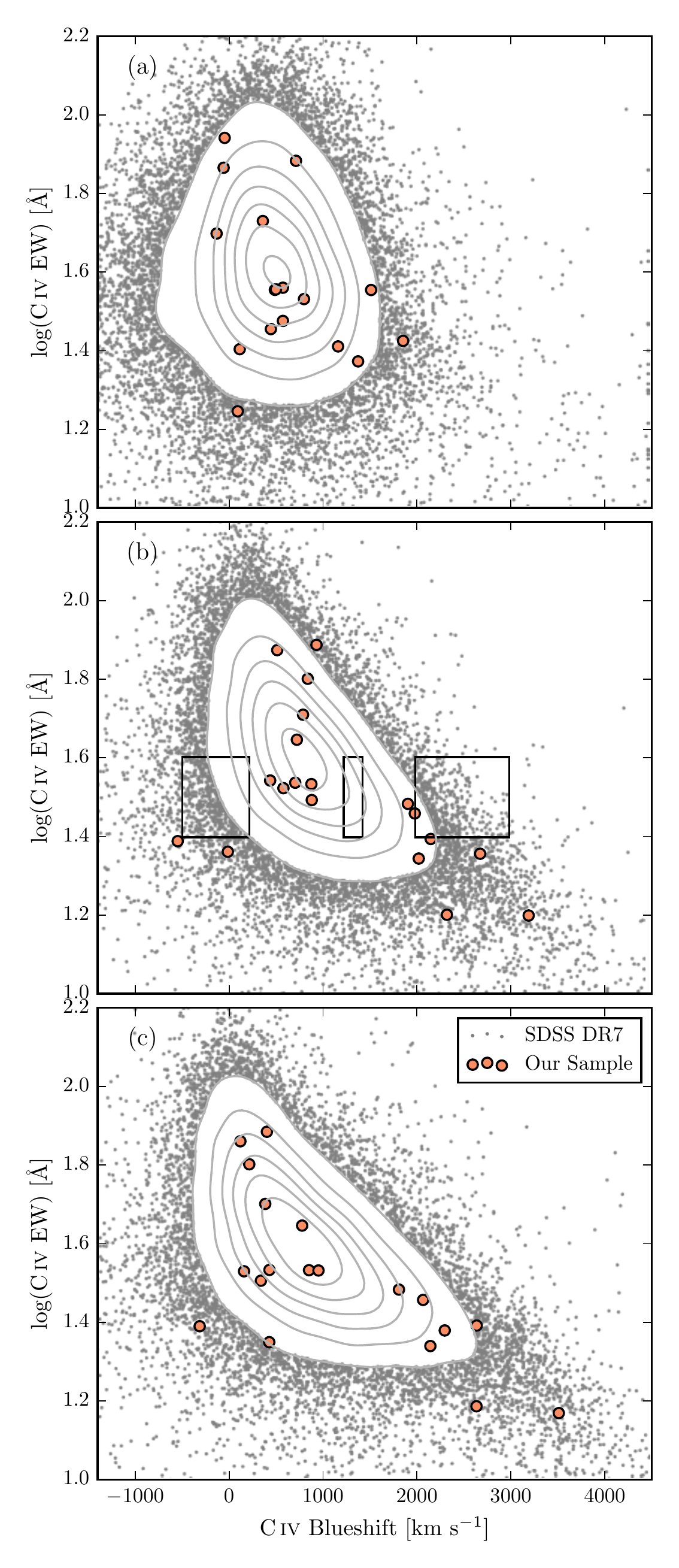}
    \caption{Rest-frame EW versus blueshift of the broad \civns-emission line for 32,157 SDSS DR7 quasars at $1.6 < z < 3.0$ ({\it grey}) and our sample ({\it orange}). Panel (a) uses \civ line parameters from \citet{shen11} and SDSS pipeline systemic redshifts. Panels (b) and (c) use systemic redshifts from \citet{hewett10} and Allen \& Hewett (2016, in preparation) respectively, and \civ line measurements described in Sec.~\ref{sec:civmeasure}. In regions of high point-density, contours show equally-spaced lines of constant probability density generated using a Gaussian kernel-density estimator. The three rectangles in panel (b) show the regions of parameter space used to generate the composite spectra shown in Fig.~\ref{fig:civ_composites}. } 
    \label{fig:civ_space}
\end{figure}

Historically, the parametrisation of the \civ emission-line properties for quasars in large surveys has not proved straightforward because the \civ emission line has itself been used in the determination of the quasar redshifts. 
The SDSS provided the first catalogue of tens of thousands of redshift $z>$1.6 quasars with spectra of adequate velocity resolution and S/N that effective statistical studies of the rest-frame ultraviolet emission-line properties, including line-shape, have proved possible.

The comprehensive compilation of quasar properties for the SDSS DR7 quasars by \citet{shen11} provides a natural starting point for population studies. 
In Fig.~\ref{fig:civ_space}a we plot the \civns-blueshift versus \civns-emission equivalent width (EW) using the SDSS pipeline redshifts and the blueshifts calculated by \citet{shen11}. 
The grey points show all SDSS DR7 quasars for which measurements exist and the orange circles show the 19 quasars with near-infrared spectra presented in this paper.  
A strong trend in the blueshift values as a function of line EW is not evident in Fig.~\ref{fig:civ_space}; structure in the parameter space is being masked because the \civ emission line is itself being used in the determination of the quasar redshifts. 

The redshift-determination scheme of \citet{hewett10} provided much improved redshifts, not least because the redshift estimates for the majority of quasars were derived using emission-lines other than the \civns-line itself. 
Figure \ref{fig:civ_space}b shows SDSS DR7 quasars in the same \civ parameter space as Figure \ref{fig:civ_space}a, but now using \citet{hewett10} redshifts. 
The improved redshift estimates are predominantly responsible for the differences seen in Fig.~\ref{fig:civ_space}a and b; the appearance in Fig.~\ref{fig:civ_space}b of the extension to high blueshift for quasars with low \civ EW is particularly evident.

The large systematic variation in the \civ emission-line profile within the population is evident from figures 11 and 12 of \citet{richards11}. 
The plots and analysis in \citet{richards11} employ the quasar redshifts from \citet{hewett10} but, as is evident from the figures, the systematic variation in the \civ shape is correlated with changes in the quasar SEDs, including the strengths of the \siiii$\lambda$1892 and \ciii$\lambda$1908 emission lines in the rest-frame ultraviolet. As a consequence, the redshifts from \citet{hewett10} still suffer from systematic errors that are correlated with the shape, and particularly the blueshift, of the \civ emission line.
The nature of the systematic variations in the quasar ultraviolet SEDs are such that for quasars with close-to symmetric \civ profiles and line centroids close to the systemic redshift, the \citet{hewett10} redshifts result in \civ blueshifts that are overestimated by a few hundred \kms, whereas, for quasars with strong blue-asymmetric \civ profiles and line centroids displaced significantly to the blue of the systemic redshift, the \civ blueshifts are underestimated by, in the most extreme cases, up to 1200\kms. 

Figure~\ref{fig:civ_space}c shows the \civ emission line parameters calculated using a new redshift-estimation algorithm (Allen \& Hewett 2016, in preparation) that takes account of the quasar SED variations, producing redshifts independent of the large systematic shape changes seen in the \civ emission line. 
The low-ionization emission lines visible in the rest-frame ultraviolet (over wavelengths from \mgii$\lambda\lambda$2796,2803 down to the \oi$\lambda$1304+\siii$\lambda$1307 blend) using the new redshift-algorithm are located at rest-frame wavelengths in excellent agreement with the systemic redshift defined using the rest-frame narrow-line optical \oiii$\lambda\lambda$4960,5008 and broad-line \hb and \hans.

The systematic trends seen in Fig.~\ref{fig:civ_space}b, in particular the extension to high blueshift at low \civ EW, become more apparent in Fig.~\ref{fig:civ_space}c, as expected from consideration of the known SED-related errors in the redshifts from \citet{hewett10}.
A population of quasars with only modest blueshifts and low EW is also apparently still present. 

\subsection{\civ emission line blueshift measurements}
\label{sec:civmeasure}

The differences in the distribution of \civ emission line properties seen in the three panels of Fig.~\ref{fig:civ_space} are due primarily to the change in the systemic redshift estimates. 
It is also necessary, however, to obtain a measure of the \civ emission line `location' in order to calculate the blueshifts. 
When working with moderately-sized samples, parametric fits to the emission-line profile may be undertaken using careful mask-definition to minimise the effect of absorption features on the profiles used for the parametrization, and this is the approach we follow below in Section~\ref{sec:line_measurements}. 
Effective analysis of the tens of thousands of spectra from SDSS DR7, and now DR12, however, requires a more robust scheme to determine a \civns-blueshift estimate that is not very sensitive to the range of S/N among the spectra or the presence of narrow absorption systems within the \civns-emission profile. 
\citet{shen11} provide a discussion (their section~3) of the factors that effect the measurement of broad emission lines in quasar spectra of modest S/N. 
Their careful analysis of the \civ emission properties employed the results of parametric fits of three Gaussians to the spectra. 
Our own experiments in quantifying the \civ emission properties of SDSS spectra showed that a simple non-parametric measure of the \civ emission location reduced the number of outliers significantly. 
Visual inspection of spectra demonstrated that the improvement is due primarily to the identification of, and interpolation over, associated and outflow absorption systems, which forms part of the non-parametric measurement scheme. 

We therefore chose to use a non-parametric scheme to measure the blueshift of the \civ line, which we will now describe. 
A continuum is first defined as a power-law of wavelength, $f(\lambda) \propto \lambda^{-\alpha}$, with the slope, $\alpha$, determined using the median\footnote{The median is used to improve the robustness of the continuum estimate from the relatively small wavelength intervals.} values of the flux in two continuum windows at 1445--1465 and 1700--1705\AA\, (the same wavelengths as adopted by \citet{shen11}). 
The \civ emission line is taken to lie within the wavelength interval 1500-1600\AA, a recipe that is commonly adopted \citep[e.g.][]{shen11, denney13}. 
To reduce the impact of narrow absorption systems on the emission-line profile a `pseudo continuum' is defined by applying a 41-pixel median filter to the quasar spectrum.
Pixels within the \civ profile that lie more than 2$\sigma$ below the pseudo-continuum are deemed to be affected by absorption and added to an `absorber'-mask. 
Two pixels on either side of each such pixel are also included in the mask. 
For each masked pixel, the flux values in the spectrum are replaced by values from the pseudo-continuum. 

The wavelength that bisects the cumulative total line flux, $\lambda_{half}$, is recorded and the blueshift (in \kms) defined as $c\times$(1549.48-$\lambda_{half}$)/1549.48 where $c$ is the velocity of light and 1549.48\AA \ is the rest-frame wavelength for the \civ doublet\footnote{The adopted \civ rest-frame wavelength assumes an optically thick BLR, in which case the contribution from each component is equal. Adopting a 2:1 ratio (appropriate for an optically thin BLR) changes the blueshifts by $\sim$80\kms.}. 
Positive blueshift values indicate an excess of emitting material moving towards the observer and hence out-flowing from the quasar.
\citet{hewett10} redshifts are used to define the quasar rest-frame. 

\subsection{Sample selection - \civ properties}

The primary aim of the paper is to investigate the potential systematic effects on the \civns-emission based BH masses for quasars with large, $\gtrsim$1200\kms, \civ blueshifts, using the properties of the \ha emission line to provide BH-mass estimates for the objects unbiased by non-virial contributions to the emission-line profile.
The orange symbols in Fig.~\ref{fig:civ_space} show the \civ parameters of our quasar targets for which near-infrared spectra of adequate S/N were obtained. 
These quasars were selected using our non-parametric blueshift measures (based on the \citet{hewett10} redshifts). 
The sample of 19 quasars spans the full dynamic range in \civns-parameters based on the \citet{hewett10} systemic redshifts and the coverage is in fact even more complete when using the forthcoming SED-independent redshifts from Allen \& Hewett (2016, in preparation).
As is evident from the sparsity of quasars with large \civ blueshifts when the SDSS pipeline systemic redshifts are used (Fig.~\ref{fig:civ_space}a), improvements in the estimation of systemic redshifts from ultraviolet spectra have been a crucial factor in allowing us to reliably select a sample of quasars with a range of \civ blueshifts. 
In subsequent sections we re-derive the systemic redshifts and \civ blueshifts for this sample using parametric fits to the \ha and \civ emission (the former from our-near infrared observations). 
Thus, while the systematic trends in BH masses inferred from measurements of the \civ emission line depend on the distribution of \civ emission line properties within the quasar population, the results of our analysis of the \ha and \civ emission line properties are independent of the redshifts used to produce the panels in Fig.~\ref{fig:civ_space}.  

\subsection{Relation to virial BH mass estimates}
\label{sec:blueshiftmasses}

\begin{figure}
    \includegraphics[width=\columnwidth]{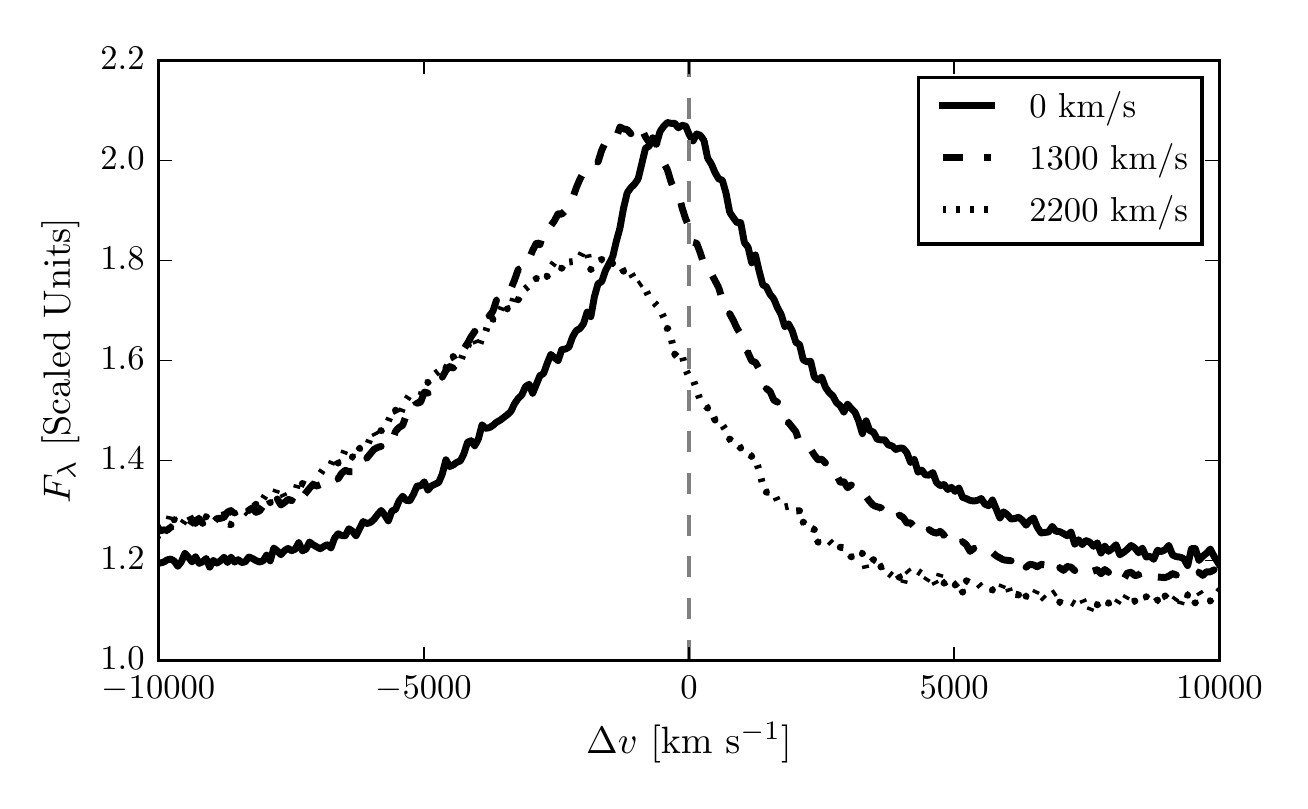}
    \caption{Composite spectra of the \civns-emission line as a function of \civ blueshift for SDSS DR7 quasars. The quasars contributing to each composite are indicated in Fig.~\ref{fig:civ_space}b. Virtually the entire \civns-profile appears to shift blueward and the change in line shape is not simply an enhancement of flux in the blue wing of a still identifiable symmetric component. In order of increasing \civ blueshift, the composite spectra have FWHM 4870, 5610, and 6770 \kms\, and EW 33.1, 31.6, and 28.8 \AA.}
    \label{fig:civ_composites}
\end{figure}

In general, researchers studying quasar demographics at high-redshift adopt estimates of BH masses based on the width of \civns-emission, without reference to the blueshift of the \civns-emission \citep[e.g.][]{vestergaard04,kollmeier06,gavignaud08,vestergaard08,vestergaard09,kelly10,kelly13}.  
The systemic redshift is often assumed to be given by the peak of the \civ emission, regardless of whether there is evidence that the line is shifted or not.
Figure~\ref{fig:civ_composites} shows the shape of the \civns-emission in composite spectra constructed from SDSS DR7 quasars with EW(\civns)=25-40\AA, as a function of \civ blueshift. 
Quasars classified as BALs, or possessing strong associated absorbers have been excluded, and the composite-spectra shown are derived using an arithmetic mean of a minimum of 200 spectra at each blueshift. 
The blueshifts and EWs of the quasars contributing to each of the composites are indicated by the boxes in Fig.~\ref{fig:civ_space}b.  
The profiles show how, at large values of blueshift ($\gtrsim$2000\kms) the \civns-profile is displaced to the blue by amounts comparable to the FWHM of the profile.

A possible origin of the blueshifts is the presence of a disc-wind \citep[see][for recent papers]{gallagher15, higginbottom15} but, irrespective of the physical origin of the high-blueshift \civns-profiles, measures of the emission-line `width' do not relate simply to virialized motions of the emitting gas under the gravitational influence of the BH. 
On the other hand, \citet{denney13} point out that any radiatively driven wind will have a velocity comparable to the escape velocity, i.e. approximately twice the virial velocity.
Even if dominated by an outflow component, the \civ line width might therefore still be expected to relate to the BH mass.

\section{Observations}
\label{sec:observations}

\begin{table*}
  \centering
  \vspace*{-0.4cm}
  \caption{Summary of near-infrared spectroscopic observations with LIRIS.}
  \label{tab:obsproperties}
  \vspace*{-0.1cm}
  \begin{minipage}{16cm}
    \begin{tabular}{ccccccccc} 
    \hline
    SDSS Name & SDSS DR & $z$\footnote{From \citet{hewett10}.} & $i_{\rm SDSS}$ & UTC Date & T$_{\rm exp}$ (s) & S/N(\civns)\footnote{\label{footnote-label}Measured in the continuum and quoted per resolution element.} & S/N(\hbns)\textsuperscript{\small \ref{footnote-label}} & S/N(\hans)\textsuperscript{\small \ref{footnote-label}} \\
    \hline
    073813.19+271038.1 & DR12 & 2.4508 & 18.80 & 2015-04-01 & 720 & 29.13 & 17.27 & 10.0 \\
    074352.61+245743.6 & DR12 & 2.1659 & 19.09 & 2015-04-04 & 2160 & 8.64 & 18.68 & 11.43 \\
    080651.54+245526.3 & DR12 & 2.1594 & 18.91 & 2015-04-01 & 1200 & 10.37 & 6.04 & 3.91 \\
	085437.59+031734.8 & DR7 & 2.2504 & 18.41 & 2015-03-31 & 2520 & 26.58 & 5.82 & 3.09 \\
	085856.00+015219.4 & DR12 & 2.1675 & 17.62 & 2015-04-04 & 1800 & 66.16 & 24.37 & 12.71 \\
	110454.73+095714.8 & DR12 & 2.4238 & 19.12 & 2015-04-03 & 1440 & 19.41 & 10.75 & 7.81 \\
	123611.21+112921.6 & DR12 & 2.1527 & 18.53 & 2015-04-04 & 1680 & 35.24 & 21.3 & 11.51 \\
	124602.04+042658.4 & DR12 & 2.4473 & 18.49 & 2015-04-01 & 960 & 35.34 & 8.1 & 5.73 \\
	130618.60+151017.9 & DR12 & 2.4020 & 19.00 & 2015-04-05 & 840 & 29.83 & 8.91 & 5.13 \\
	131749.78+080616.2 & DR12 & 2.3791 & 19.04 & 2015-04-05 & 2880 & 19.25 & 5.6 & 3.32 \\
	132948.73+324124.4 & DR12 & 2.1684 & 18.40 & 2015-04-01 & 2520 & 32.58 & 10.4 & 6.96 \\
	133646.87+144334.2 & DR7 & 2.1422 & 18.84 & 2015-04-01 & 1200 & 15.2 & 23.82 & 16.34 \\
	133916.88+151507.6 & DR12 & 2.3157 & 18.52 & 2015-04-03 & 2880 & 20.52 & 5.79 & 3.28 \\
	140047.45+120504.6 & DR12 & 2.1722 & 18.29 & 2015-04-02 & 840 & 36.64 & 9.83 & 5.68 \\
	152529.17+292813.2 & DR12 & 2.3605 & 17.52 & 2015-04-04 & 1440 & 80.55 & 2.17 & 1.5 \\
	153027.37+062330.8 & DR12 & 2.2198 & 18.62 & 2015-04-04 & 1800 & 29.9 & 21.01 & 12.58 \\
	153848.64+023341.1 & DR12 & 2.2419 & 17.56 & 2015-04-01 & 2520 & 64.82 & 5.63 & 3.56 \\
	161842.44+234131.7 & DR7 & 2.2824 & 18.49 & 2015-04-04 & 1320 & 23.37 & 11.1 & 6.43 \\
	163456.15+301437.8 & DR12 & 2.4901 & 18.29 & 2015-04-01 & 1920 & 36.06 & 9.44 & 8.63 \\
    \hline
    \end{tabular}
  \vspace*{-0.4cm}
  \end{minipage}
\end{table*}

Near-infrared spectra were obtained with the Long-slit Intermediate Resolution Infrared Spectrograph (LIRIS) mounted on the 4.2m William Herschel Telescope (WHT) at the Observatorio del Roque de los Muchachos (La Palma, Spain). 
Observations took place over four non-contiguous nights from 2015 March 31 to April 4. 
Approximately one night was lost due to poor weather and a further half-night was affected by poor transparency due to cloud. 
A one arcsecond slit-width was employed and the LIRIS $H+K$ low-resolution grism was selected, which covers the spectral ranges 1.53--1.79\,$\mu$m and 2.07--2.44\,$\mu$m with a dispersion of 9.7\AA/pixel. 
The spatial scale of the instrument is 0.25 arcsec/pixel. 
Observations were divided into 60\,s sub-exposures and performed in an ABBA nodding pattern, with the object placed at two positions along the slit 12 arcsec apart. 
Bright A0-5V stars were observed at similar air-masses to the targets in order to provide both telluric absorption corrections and a flux calibration of the quasar spectra.

The raw LIRIS data frames incorporate a known `pixel shift' which was first removed from all frames using the LIRIS data reduction package {\tt LIRISDR}. 
Subsequent data reduction was undertaken with standard {\tt IRAF}\footnote{IRAF is distributed by the National Optical Astronomy Observatory, which is operated by the Association of Universities for Research in Astronomy (AURA) under a cooperative agreement with the National Science Foundation.} procedures.  
The flat-field images, which were taken at the beginning of each night via illumination of the dome, were averaged and normalised to remove any wavelength-dependent signature. 
Each individual two-dimensional spectrum was then flat-field corrected. 
Consecutive AB and BA pairs of two-dimensional spectra were subtracted to remove the sky background. 
All the subtracted AB/BA-pairs for a target were then averaged to give the final two-dimensional spectrum.

The size of the one-dimensional spectrum extraction windows, in the slit direction, varied from 6-10 pixels. 
To increase the S/N, optimal variance-weighted extraction with sigma clipping was employed. 
For the fainter objects in our sample we were unable to trace the spectrum across the dispersion axis reliably and the trace from a telluric standard-star observation, observed at a similar air mass and time, was used instead. 
The wavelength calibration, using argon and xenon lamp exposures, resulted in root mean square errors in the range 1.01--1.71\,\AA, with a mean of 1.47\AA. 
The telluric standard star observations were reduced using the same steps described above. 
The stellar continuum was divided out of the standard star spectrum, which was then divided into the quasar spectrum to remove telluric absorption features. 
The spectral type and magnitude of the standard star were used to flux calibrate the quasar spectrum both in a relative and absolute sense.
Variable atmospheric conditions combined with the narrow slit width resulted in a significant level of uncertainty in the absolute flux calibration for the quasar observations. 
The use of the UKIDSS broadband magnitudes ($H$ and $K$) to normalise the spectra results in a significantly improved calibration. 

\begin{figure}
    \includegraphics[width=\columnwidth]{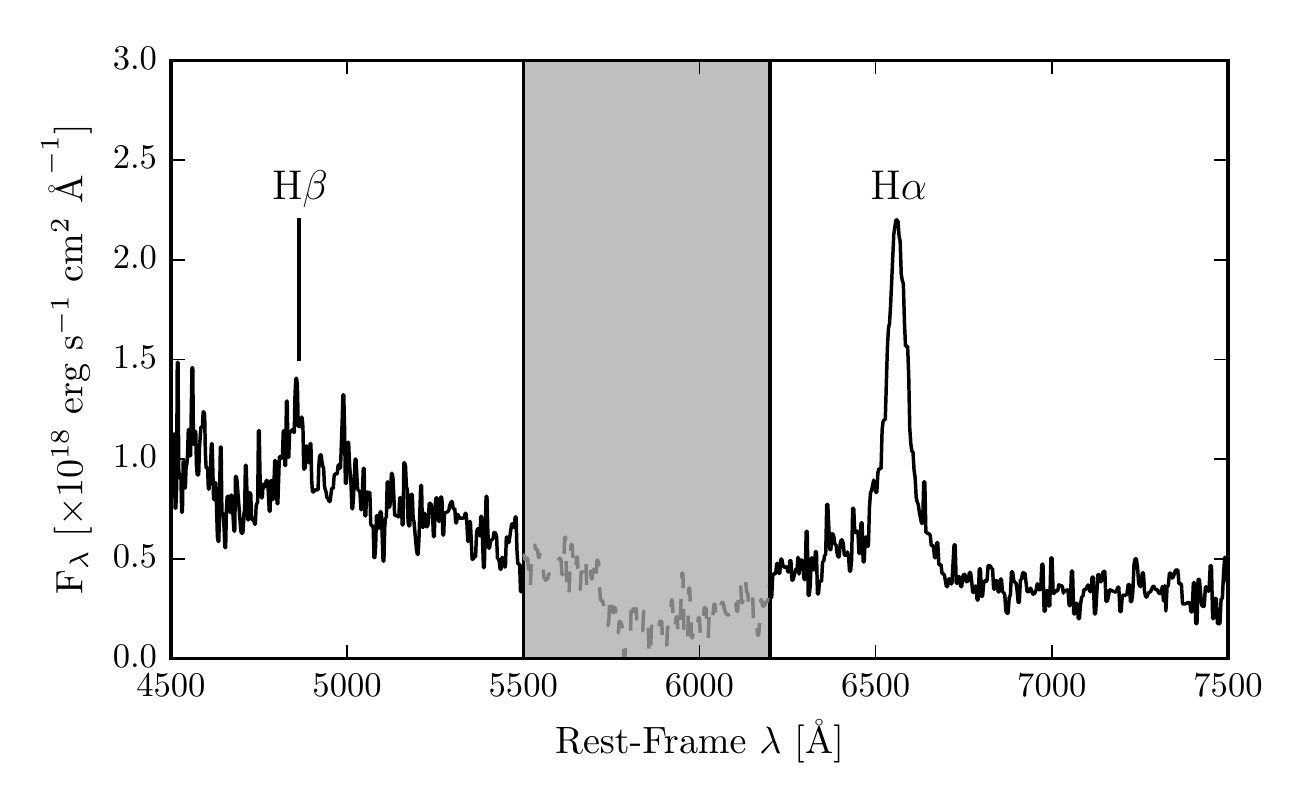} 
    \caption{LIRIS spectrum of SDSSJ1246+0426. The gap between the $H$ and $K$ bands ($\sim$5500-6300\,\AA) has been greyed-out.}
    \label{fig:example_spectrum}
\end{figure}

Spectra with sufficient S/N for analysis of the \ha emission line properties were obtained for a total of 19 quasars. 
Sixteen of the 19 quasars have been re-observed in the Sloan Digital Sky Survey-III: Baryon Oscillation Spectroscopic Survey \citep[SDSS-III/BOSS;][]{dawson13} and the spectra are available in the SDSS DR12 quasar catalogue \citep{paris16}. 
As the BOSS-spectra have higher S/N compared to those in DR7, we have used the BOSS spectra when available.
A typical reduced LIRIS spectrum is shown in Fig.~\ref{fig:example_spectrum}. 
A log of the observations, the quasar positions, magnitudes and redshifts, along with the S/N achieved for the \hbns, \ha and \civ emission line regions (the last from the optical SDSS/BOSS spectra) are listed in Table~\ref{tab:obsproperties}.
The S/N, which is given per resolution element, was measured in the continuum in the region around the emission lines. 
The full SDSS name is given in Table~\ref{tab:obsproperties}; in the subsequent tables and text we will refer to objects using an abbreviated name of the form SDSSJHHMM+DDMM.

Although the S/N is similar in the continuum regions adjacent to the \ha and \hb emission lines, in practice the much lower EW of \hb compared to \ha meant that both parametric and non-parametric characterisation of the emission-line parameters did not produce results that could be used in this investigation. 
The individual \hb profiles were thus not employed, although a composite spectrum of the \hb region is used below.

\section{Emission line measurements} 
\label{sec:line_measurements}

Virial BH mass estimators are calibrated using either the FWHM or dispersion ($\sigma$; derived from the second-moment velocity) of a broad emission line \citep[e.g.][]{vestergaard06,park13}. 
Complications which are encountered when measuring line widths include how to model the `continuum' flux, where to define the limits of the line emission, and how to deal with absorption. 
All of these issues are exacerbated when working with low S/N data \citep[see][for a discussion]{denney13}. 
In Section~\ref{sec:civmeasure} we measured the blueshift of \civ for tens of thousands of SDSS DR7 quasar spectra. 
This allowed us to quantify the distribution of \civ blueshift values and hence select a subset for near-infrared observations which have \civ blueshifts spanning the full range of this distribution (Fig.~\ref{fig:civ_space}).
A non-parametric scheme was employed because, in comparison to recipes involving the fitting of multiple Gaussian (or other parametric) profiles, it was found to be more robust and less sensitive to the range of S/N among the spectra and to the presence of narrow absorption systems within the \civns-emission profile.
In this section we will use a different approach, and measure the line properties by fitting a parametric model to the data. 
When working with a small number of spectra, it is possible to use careful mask-definition to minimise the effect of absorption features on the profiles used for the parametrization.
The purpose of the model fits is purely to best represent the intrinsic line profile, and no physical meaning is attached to the individual model components. 
We will now describe the parametric model and fitting procedure used for each emission line. 
The models were fit using a standard variance-weighted least squares minimisation procedure employing the Levenberg-Marquardt algorithm. 
Prior to the fit, the spectra were visually inspected and regions significantly affected by absorption were masked and excluded.

\begin{table*}
  \centering
  \caption{Summary of the fitting regions and the parameters of the models used to fit the \civ and \ha emission lines.}
  \label{tab:fittingproperties}
    \begin{tabular}{ccccccccc}
    \hline
    & \multicolumn{2}{c}{Fitting Region [\kms]} & \multicolumn{2}{c}{Continuum Region[\AA]} & GH Order & Gaussians & \multicolumn{2}{c}{$\chi^2_{\nu}$} \\
    Name & \civns  & \hans & \civns & \hans & \civns & \hans & \civns & \hans \\
    \hline
    0738+2710 & -9570,9770 & -7530,10740 & 1445-1465,~~1700-1705 & 6000-6250,~~6800-7000 & 6 & 2 & 0.66 & 1.0 \\
    0743+2457 & -9570,9770 & -7530,10740 & 1445-1465,~~1700-1705 & 6004-6210,~~6800-7000 & 2 & 2 & 0.84 & 1.0 \\
    0806+2455 & -9570,9770 & -9219,10759 & 1445-1465,~~1700-1705 & 6000-6250,~~6800-7000 & 3 & 1 & 0.87 & 0.83 \\
    0854+0317 & -9570,9770 & -7530,10740 & 1445-1465,~~1700-1705 & 5989-6135,~~6800-7000 & 3 & 2 & 0.91 & 0.85 \\
    0858+0152 & -20000,7400 & -7530,10740 & 1423-1428,~~1700-1705 & 6000-6200,~~6800-7000 & 2 & 2 & 0.94 & 0.96 \\
    1104+0957 & -9570,9770 & -7530,10740 & 1445-1465,~~1700-1705 & 6000-6250,~~6801-6845 & 6 & 2 & 0.68 & 0.95 \\
    1236+1129 & -15363,7650 & -8904,10590 & 1445-1465,~~1700-1705 & 6063-6210,~~6800-7000 & 3 & 2 & 0.84 & 0.89 \\
    1246+0426 & -9570,9770 & -7530,10740 & 1445-1465,~~1700-1705 & 6000-6250,~~6799-6906 & 4 & 2 & 0.66 & 0.83 \\
    1306+1510 & -13000,7800 & -7530,10740 & 1445-1465,~~1700-1705 & 6000-6250,~~6800-7000 & 2 & 3 & 0.73 & 0.18 \\
    1317+0806 & -9198,9755 & -7530,10740 & 1445-1465,~~1700-1705 & 6000-6250,~~6800-7000 & 2 & 1 & 0.78 & 2.16 \\
    1329+3241 & -12000,9000 & -7605,7406 & 1445-1461,~~1700-1705 & 6000-6250,~~6800-7000 & 3 & 2 & 0.61 & 0.84 \\
    1336+1443 & -14000,10000 & -10131,10674 & 1445-1465,~~1700-1705 & 6000-6250,~~6800-7000 & 3 & 2 & 0.87 & 1.51 \\
    1339+1515 & -12000,11000 & -7530,10740 & 1445-1465,~~1700-1705 & 6046-6250,~~6800-7000 & 4 & 1 & 0.69 & 0.14 \\
    1400+1205 & -15000,10000 & -2000,10815 & 1445-1465,~~1700-1705 & 6000-6250,~~6800-7000 & 6 & 2 & 0.82 & 0.2 \\
    1525+2928 & -9570,9770 & -7586,8080 & 1459-1466,~~1700-1705 & 6055-6251,~~6800-7000 & 4 & 1 & 0.49 & 0.39 \\
    1530+0623 & -12000,10000 & -7530,10740 & 1445-1465,~~1700-1705 & 6127-6186,~~6800-7000 & 4 & 3 & 0.84 & 1.14 \\
    1538+0233 & -13500,9000 & -7530,10740 & 1450-1465,~~1700-1705 & 6000-6250,~~6855-7002 & 4 & 2 & 0.61 & 0.82 \\
    1618+2341 & -9190,9770 & -7530,10740 & 1445-1465,~~1689-1697 & 6000-6250,~~6800-7000 & 4 & 2 & 1.16 & 0.93 \\
    1634+3014 & -9570,9770 & -8400,8500 & 1445-1465,~~1700-1705 & 6000-6250,~~6736-6779 & 3 & 2 & 0.59 & 0.22 \\
    \hline
  \end{tabular}
\end{table*}

\subsection{\civns}

We first measure and subtract the local continuum emission, by fitting a power-law to two windows on either side of the line emission, as described in Section~\ref{sec:civmeasure}. 
For a small number of objects, absorption features, or artefacts, in the spectrum necessitated modest adjustments to the window extents, which are specified in Table~\ref{tab:fittingproperties}. 
The continuum-subtracted spectra are then transformed from wavelength units into units of velocity relative to the rest-frame line-transition wavelength for the \civ doublet (1549.48\AA, assuming equal contributions from both components). 
The parametric model is ordinarily fit within the same 1500--1600\,\AA \, window used in Section~\ref{sec:civmeasure}, which corresponds to approximately $\pm 10\,000$ \kms\, from the rest-frame transition wavelength. 
The line-window was extended if significant flux in the profile was present blueward of the short wavelength limit. 
The adopted line-fitting windows, in units of velocity from the rest-frame transition wavelength, are given in Table~\ref{tab:fittingproperties}. 

\begin{figure}
    \includegraphics[width=\columnwidth]{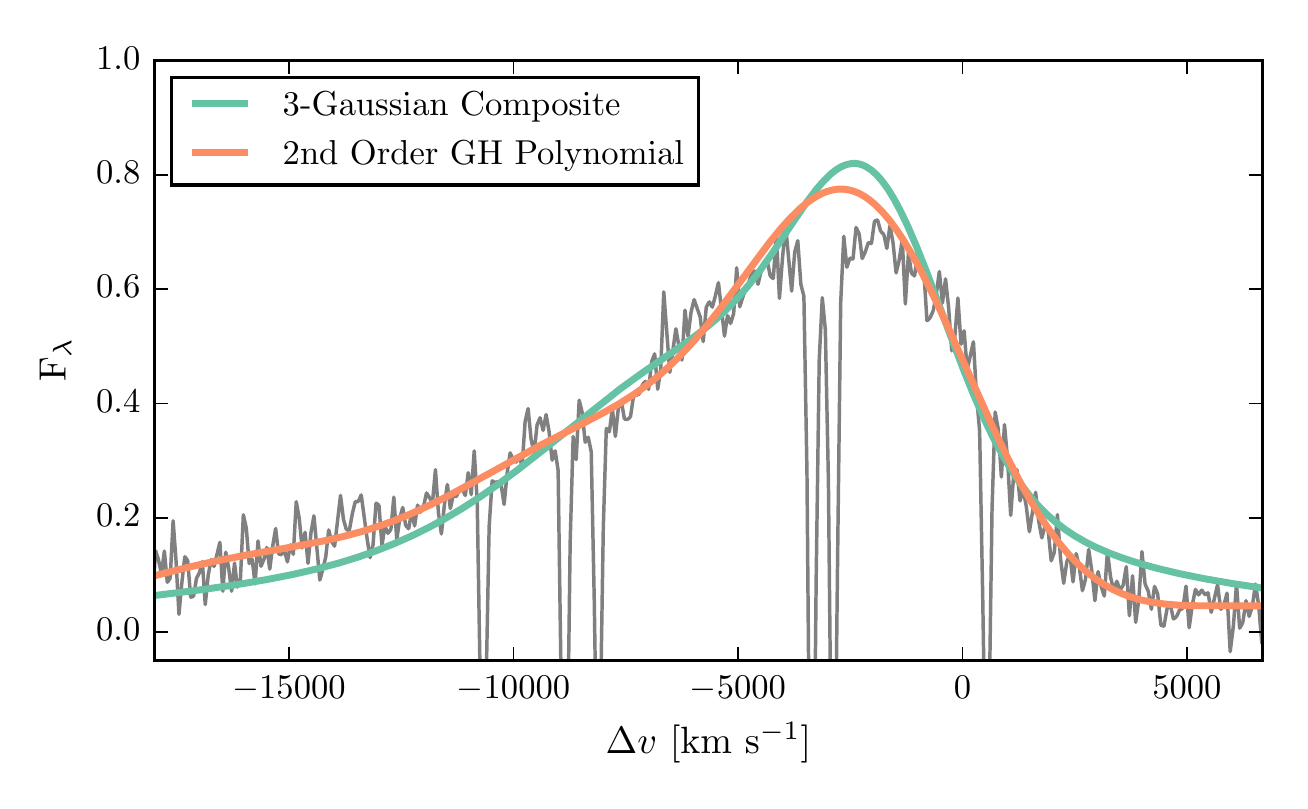} 
    \caption{2nd-order Gauss-Hermite (GH) polynomial and 3-component Gaussian fit to the \civns-emission line of SDSSJ0858+0152, which is the most blueshifted in our sample. We derive line parameters from the GH polynomial fit; using the Gaussian model changes the FWHM, line dispersion, and blueshift by -250, -150, and 500 \kms respectively. For the \civ lines of all other quasars in our sample the GH polynomial and Gaussian models provide equally good fits.} 
    \label{fig:fitting_comparison}
\end{figure}

To fit the \civ profile we employed Gauss-Hermite (GH) polynomials, using the normalisation of \citet{marel93} and the functional forms of \citet{cappellari02}.
We allowed up to six components in the GH polynomial model, but in many cases a lower order was sufficient; the polynomial order used for each line is given in Table~\ref{tab:fittingproperties}.
It is also a common practice to fit the \civ emission profile with two or three Gaussian components \citep[e.g.][]{shen11}. 
We opted to use a GH-polynomial model primarily because it provided a significantly better fit to the most blueshifted and asymmetric \civ line (in SDSSJ0858+0152).  
Figure~\ref{fig:fitting_comparison} shows how a model with three Gaussian components underestimates the flux in the blue wing and overestimates the flux in the red wing of the line profile. 
Using the Gaussian model rather than the GH polynomial changes the FWHM, line dispersion, and blueshift by -3, -3, and 10\,per cent respectively.
We have highlighted SDSSJ0858+0152 because, of all the objects in our sample, the choice of model leads to the largest change in \civ line parameters.
Even in this case, however, the differences are modest. 

For every other \civ line in our sample we found only marginal differences in our best-fit line parameters when, rather than using a GH polynomial model, the \civ emission was fit using a composite model of up to three Gaussians. 
Our best-fit parameters are also in good agreement with \citet{shen11}, who employ a multi-Gaussian parametrization\footnote{The \citet{shen11} parameters are derived from the SDSS DR7 spectra, whereas 16 out of 19 of our fits are to higher S/N BOSS DR12 spectra.}.  
The scatter between the \citet{shen11} results and our own is 0.1\,dex about the one-to-one relation and, as expected, is larger for lines with smaller EWs. 

\subsection{\ha}

We employ the same continuum subtraction and fitting method as for \civns, with the continuum and fitting windows as given in Table~\ref{tab:fittingproperties}. 
We adopt a rest-frame transition wavelength of 6564.89\,\AA\, to transform wavelengths into equivalent Doppler velocities. 
We used a simple model with up to three broad Gaussian components to fit the \ha emission line.
We opted against parametrizing the \ha line using a GH polynomial because the extra degrees of freedom in this model did not improve the quality of the fits\footnote{The emission line parameters and subsequent analysis do not depend on whether line parameters from multiple-Gaussian or GH-polynomial model fits are used.}.
Upon inspection of the residuals from the fit, we also found no evidence that additional model components for narrow \hans, \niins\ll6548,6584 and \siins\ll6717,6731 were required.
Furthermore, narrow \oiiins\ll4960,5008 emission is relatively weak in these spectra.

The sole exception is the \ha line in the spectrum of SDSSJ0738+2710.
In addition to having the narrowest \ha line, this spectra also has the strongest narrow \oiii component (EW = 63\,\AA), which suggests that a contribution from the narrow-line region might be important.  
Introducing a single Gaussian for the narrow emission, while retaining a double Gaussian for the broad emission, the FWHM of the broad component increases to 3400\,\kms (compared to 1580\,\kms without the narrow component). 
For consistency, the parameters quoted in Table~\ref{tab:emissionproperties} are from the model with no narrow component. 
However, because the properties derived from the emission line width (the BH mass and the mass-normalised accretion rate) are strongly biased by the probable contribution from the narrow-line region, SDSSJ0738+2710 is excluded from the analysis in Section~\ref{sec:results}. 

\subsection{Comparison of H$\alpha$ and H$\beta$ profiles}

Virial BH mass estimators are typically based on the width of \hbns. 
However, the \ha and \hb emission is believed to originate from the same gas and the transformation between the emission-line velocity widths is expected to be well defined. 
\citet{greene05}, using a sample of 162 quasars with high S/N SDSS spectra, established the following relation between the \ha and \hb FWHM:

\begin{equation}
  \label{eq:hb2hawidth}
  \rm{FWHM}(\rm{H}\beta) = (1.07 \pm 0.07) \times 10^3 \left( \frac{ \rm{FWHM}(\rm{H}\alpha) }{10^3 ~\rm{km}~\rm{s}^{-1} } \right)^{(1.03 \pm 0.03)}
\end{equation}

\begin{figure}
    \includegraphics[width=\columnwidth]{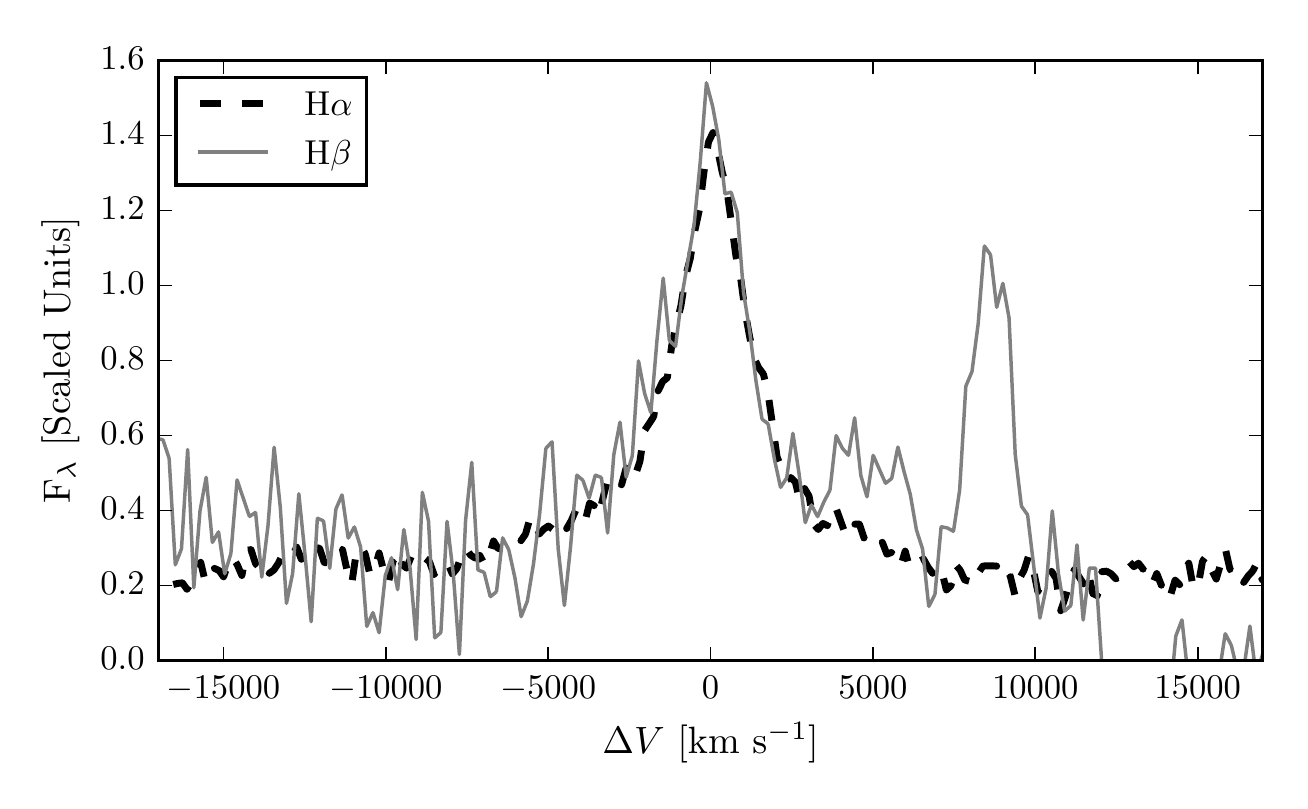} 
    \caption{The \ha and \hb emission line regions in the median composite spectrum, shown as function of the velocity shift from the respective predicted line peak wavelengths. The line fluxes have been scaled in order for the profile shapes to be readily compared. The \ha and \hb line profiles are very similar, which suggests a tight correlation between the \ha and \hb line widths. Quasar narrow-line emission from \oiii\l5008.2 is visible but, overall, the \oiii\ll4960,5008 emission is relatively weak in these spectra.}
    \label{fig:balmer_composite}
\end{figure}

\citet{greene05} found the root-mean-square scatter about this relation to be $\sim$ 0.1 dex. 
We do not have a sufficient number of robust \hb line measurements to test this relation directly.
However, we are in the process of acquiring a much larger sample of quasars with near-infrared spectra covering \ha and \hb at similar redshifts and luminosities to the sample presented here. 
The \ha and \hb line widths of this sample are in excellent agreement with the \citet{greene05} relation. 
To indirectly test the \hans/\hb line width relation for the sample presented here, we first constructed mean composite spectra in the \ha and \hb emission line regions to increase the S/N. 
The individual rest-frame spectra (defined using the wavelength of the \ha centroid) were interpolated on to a common wavelength grid. 
The spectra were then normalised using the continuum flux under the line centre, which was found by linearly interpolating between two emission-line free windows on either side of the line. 
Figure.~\ref{fig:balmer_composite} shows the composite \ha and \hb line regions overlaid and, as expected, the line profiles are closely matched.

\subsection{Emission line parameters}

\begin{table*}
  \centering
  \caption{Summary of emission line properties derived from parametric model fits to \ha and \civns. }
  \label{tab:emissionproperties}
  \begin{tabular}{ccccccccc}
  \hline
  & & Blueshift & \multicolumn{2}{c}{FWHM} & \multicolumn{2}{c}{$\sigma$} & \multicolumn{2}{c}{EW}  \\
  & & [\kms] & \multicolumn{2}{c}{[\kms]} & \multicolumn{2}{c}{[\kms]} & \multicolumn{2}{c}{[\AA]} \\
  Name & $z$ (\hans) & \civns & \civns & \hans & \civns & \hans & \civns & \hans \\
  \hline
  0738+2710 & 2.4396 & $50\pm21$ & $2255\pm42$ & $1503\pm95$ & $2916\pm42$ & $1789\pm95$ & $54\pm2$ & $532\pm43$ \\
  0743+2457 & 2.1662 & $692\pm210$ & $5924\pm536$ & $6036\pm511$ & $3801\pm536$ & $3904\pm511$ & $33\pm4$ & $351\pm23$ \\
  0806+2455 & 2.1542 & $389\pm115$ & $3435\pm267$ & $4059\pm403$ & $3387\pm267$ & $1724\pm403$ & $51\pm4$ & $488\pm56$ \\
  0854+0317 & 2.2475 & $-403\pm134$ & $3940\pm354$ & $4436\pm522$ & $3448\pm354$ & $3146\pm522$ & $24\pm2$ & $617\pm90$ \\
  0858+0152 & 2.1692 & $4354\pm82$ & $8412\pm384$ & $3155\pm80$ & $5298\pm384$ & $3758\pm80$ & $28\pm1$ & $622\pm23$ \\
  1104+0957 & 2.4217 & $-299\pm55$ & $3590\pm112$ & $3307\pm334$ & $3259\pm112$ & $2723\pm334$ & $70\pm3$ & $441\pm72$ \\
  1236+1129 & 2.1559 & $2828\pm99$ & $7540\pm271$ & $3152\pm131$ & $4168\pm271$ & $3277\pm131$ & $28\pm1$ & $631\pm29$ \\
  1246+0426 & 2.4393 & $325\pm145$ & $4126\pm88$ & $4268\pm472$ & $3901\pm88$ & $2543\pm472$ & $48\pm1$ & $536\pm77$ \\
  1306+1510 & 2.3989 & $2043\pm84$ & $6660\pm158$ & $2626\pm330$ & $3905\pm158$ & $2145\pm330$ & $36\pm1$ & $349\pm52$ \\
  1317+0806 & 2.3748 & $437\pm289$ & $5256\pm182$ & $7188\pm946$ & $3675\pm182$ & $3033\pm946$ & $33\pm2$ & $374\pm67$ \\
  1329+3241 & 2.1637 & $652\pm113$ & $4528\pm491$ & $4908\pm410$ & $3819\pm491$ & $3350\pm410$ & $35\pm2$ & $428\pm35$ \\
  1336+1443 & 2.1466 & $3668\pm345$ & $8780\pm1003$ & $2954\pm67$ & $3772\pm1003$ & $3227\pm67$ & $20\pm2$ & $523\pm17$ \\
  1339+1515 & 2.3207 & $133\pm184$ & $3865\pm935$ & $8816\pm1072$ & $4501\pm935$ & $3627\pm1072$ & $44\pm2$ & $500\pm98$ \\
  1400+1205 & 2.1672 & $2492\pm107$ & $7590\pm290$ & $3231\pm227$ & $4363\pm290$ & $3103\pm227$ & $25\pm1$ & $642\pm55$ \\
  1525+2928 & 2.3572 & $612\pm536$ & $5697\pm128$ & $6360\pm1915$ & $4303\pm128$ & $2696\pm1915$ & $41\pm1$ & $458\pm186$ \\
  1530+0623 & 2.2169 & $1471\pm108$ & $5397\pm302$ & $3073\pm145$ & $4092\pm302$ & $2664\pm145$ & $26\pm1$ & $499\pm28$ \\
  1538+0233 & 2.2420 & $2018\pm80$ & $5567\pm100$ & $2892\pm253$ & $3596\pm100$ & $2415\pm253$ & $25\pm1$ & $465\pm75$ \\
  1618+2341 & 2.2755 & $42\pm53$ & $2516\pm161$ & $2669\pm175$ & $3312\pm161$ & $2359\pm175$ & $34\pm2$ & $425\pm39$ \\
  1634+3014 & 2.5018 & $1509\pm223$ & $6835\pm745$ & $6210\pm900$ & $4566\pm745$ & $3236\pm900$ & $26\pm1$ & $327\pm65$ \\
  \hline
  \end{tabular}
\end{table*}

\begin{figure*}
	\includegraphics[width=2\columnwidth]{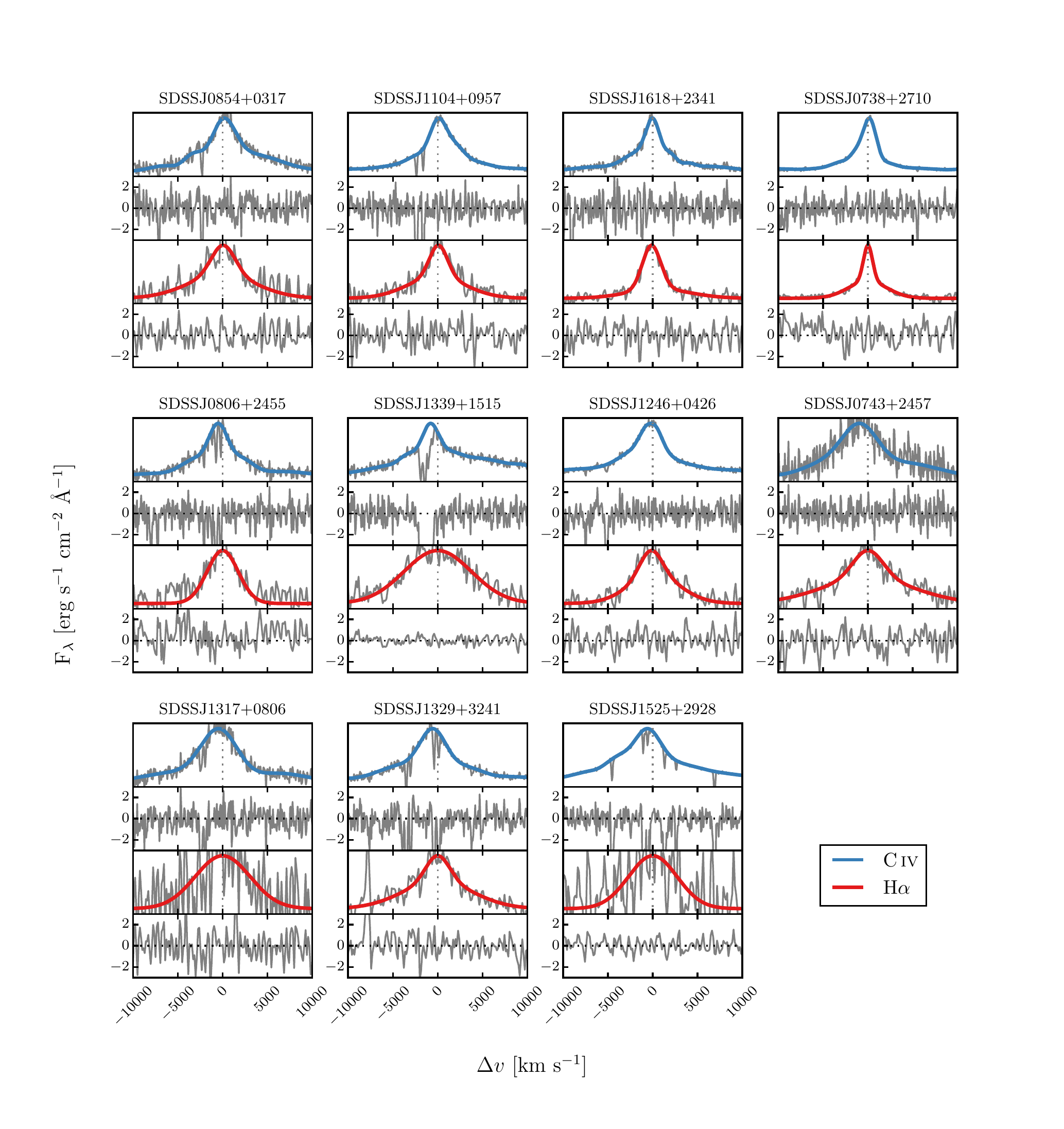} 
    \caption{\civ (SDSS/BOSS) and \ha (LIRIS) emission lines and best-fitting model. $\Delta{v}$ is the velocity shift from the line rest-frame transition wavelength, with the systemic redshift defined using the centroid of the fit to \hans. Objects are presented in order of increasing \civ blueshift relative to the \ha centroid. Below each fit we plot the data-model residuals, scaled by the errors on the fluxes. }
    \label{fig:gridspectra_1}
\end{figure*}

\renewcommand{\thefigure}{\arabic{figure}}
\addtocounter{figure}{-1}
\begin{figure*}
	\includegraphics[width=2\columnwidth]{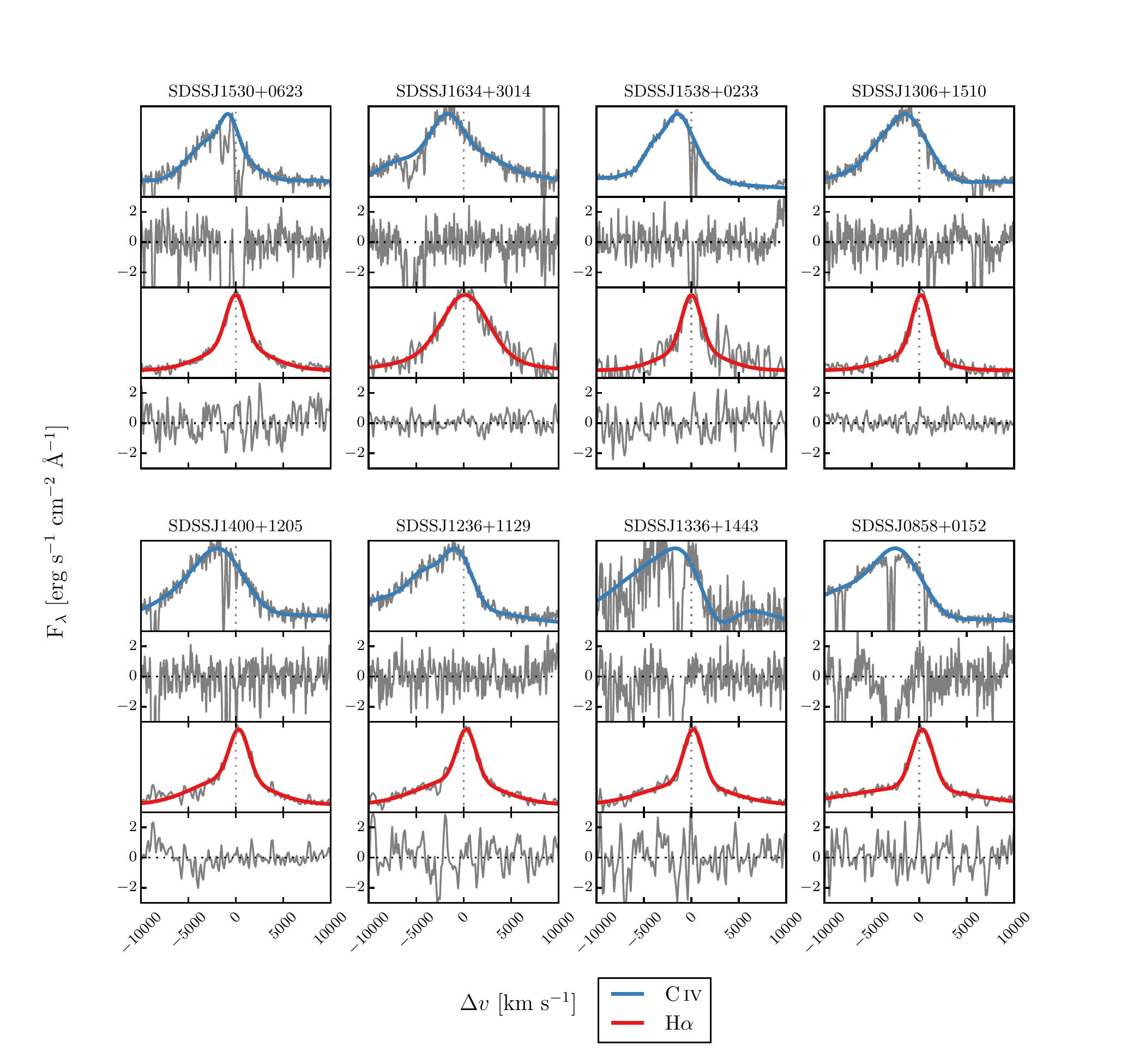}
    \caption{\it Continued}
    \label{fig:gridspectra_2}
\end{figure*}
\renewcommand{\thefigure}{\arabic{figure}}

In Fig.~\ref{fig:gridspectra_1} we show our best-fitting models overlaid on the observed flux in the spectral regions around \civ and \hans. 
The spectra are presented in order of increasing \civ blueshift. 
The Doppler velocities have been shifted so that the \ha emission line centroid is at 0\,\kms. 
The $y$-axes of the data-minus-model residual plots have deliberately been scaled by the spectrum flux errors.
The model fits are generally very good with only minimal systematic residuals. 
The only significant features seen in the residual \civ spectra correspond to the location of narrow absorption lines which were excluded in the fitting procedure.
The continuum windows for a number of the \ha lines extend close to the edges of the $K$-band and uncertainties in the flux calibration, and hence continuum level, are almost certainly responsible for the low-amplitude, large-scale, systematic residuals seen in a number of objects (e.g. SDSSJ 0806+3455, 0858+0152, 1400+1205, 1538+0233). 
The amplitudes are, however, small and redefining the continuum levels to eliminate the residuals has only a very small effect on the line-profile parameters used in the analysis. 

The systemic redshift (defined using the \ha peak), the \civ blueshift, the line FWHM, dispersion, and EW of \ha and \civ are all given in Table~\ref{tab:emissionproperties}.
Both the FHWM and the dispersion ($\sigma$) have been corrected for instrumental broadening by subtracting the FWHM resolution (152 and 477\,\kms\, for SDSS/BOSS and LIRIS respectively) in quadrature\footnote{For the dispersion we first divide the FHWM resolution by 2.35, which assumes that the line profile is Gaussian.}. 
 
Our definition of the \civ blueshift differs slightly in two ways from the values plotted for the quasar population in Fig.~\ref{fig:civ_space}. 
Firstly, we use the peak of our parametric model fit to the \ha line to define the quasar systemic-redshift\footnote{The \hans-derived redshifts are very closely in agreement with those from the forthcoming Allen \& Hewett redshifts, which are plotted in Fig.~\ref{fig:civ_space}c.}.  
Secondly, the centre of the \civ line is now defined as the wavelength that bisects the cumulative total flux of our best-fit GH-polynomial model rather than of the data. 

The monochromatic luminosity of the continuum at 1350 and 5100\,\AA, which are used to calculate virial BH mass estimates, are given in Table~\ref{tab:continuumproperties}. 
The luminosity at 1350\,\AA\, was taken from the spectral fits of \citet{shen11}.
The quasar rest-frame continuum at 5100\AA \ often lies at the edge, or beyond, the wavelength coverage of the LIRIS spectra. 
Monochromatic 5100\AA \ luminosities were therefore calculated from the fit of our parametric quasar model (described in \citet{maddox12}) to the UKIDSS $H$- broadband magnitude for each quasar. 
The model fits to the quasars are excellent, with residuals in SDSS and UKIDSS passbands under 10\,per cent. 

\begin{table}
  \centering
  \caption{Monochromatic continuum luminosities used to derive bolometric luminosities and BH masses.}
  \label{tab:continuumproperties}
  \begin{tabular}{ccc}
  \hline
  & \multicolumn{2}{c}{Log $L_\lambda$} \\
  & \multicolumn{2}{c}{[erg~s$^{-1}$]} \\
  Name & 1350\AA & 5100\AA \\
  \hline
  0738+2710 & $46.43\pm0.02$ & $46.14\pm0.01$ \\
  0743+2457 & $46.07\pm0.02$ & $45.93\pm0.02$ \\
  0806+2455 & $46.09\pm0.01$ & $45.95\pm0.02$ \\
  0854+0317 & $46.28\pm0.01$ & $46.27\pm0.01$ \\
  0858+0152 & $46.82\pm0.00$ & $46.37\pm0.01$ \\
  1104+0957 & $46.11\pm0.03$ & $45.92\pm0.02$ \\
  1236+1129 & $46.45\pm0.01$ & $46.01\pm0.01$ \\
  1246+0426 & $46.46\pm0.01$ & $46.13\pm0.01$ \\
  1306+1510 & $46.35\pm0.01$ & $46.00\pm0.02$ \\
  1317+0806 & $46.31\pm0.01$ & $46.02\pm0.01$ \\
  1329+3241 & $46.35\pm0.01$ & $46.08\pm0.02$ \\
  1336+1443 & $45.84\pm0.02$ & $45.95\pm0.01$ \\
  1339+1515 & $46.42\pm0.01$ & $45.97\pm0.01$ \\
  1400+1205 & $46.45\pm0.01$ & $46.05\pm0.01$ \\
  1525+2928 & $46.84\pm0.01$ & $46.50\pm0.01$ \\
  1530+0623 & $46.26\pm0.01$ & $45.97\pm0.01$ \\
  1538+0233 & $46.94\pm0.00$ & $46.51\pm0.01$ \\
  1618+2341 & $46.59\pm0.01$ & $46.10\pm0.01$ \\
  1634+3014 & $46.66\pm0.01$ & $46.16\pm0.01$ \\
  \hline
  \end{tabular}
\end{table}

\subsection{Emission line parameter uncertainties}
\label{sec:param_uncertainties}

The 1$\sigma$ error bars calculated from the covariance matrix in least-squares minimisation will underestimate the true uncertainties on the line parameters, since they do not account for systematic errors such as the significant uncertainty introduced in the continuum subtraction procedure.  
To calculate more realistic uncertainties on our fitted variables we employed a Monte Carlo approach. 
Artificial spectra were synthesised, with the flux at each wavelength drawn from a Normal distribution (mean equal to the measured flux and standard deviation equal to the known error). 
Our emission-line fitting recipe was then implemented on five thousand artificial spectra.  
Our parameter uncertainties are defined as the standard deviation of the best-fitting parameter values from these five thousand realisations.  
The uncertainty on the monochromatic continuum luminosity at 5100\,\AA\, was estimated via a very similar method -- using the error on the UKIDSS $H$-band magnitude to run a number of realisations of our SED-fitting routine. 
The uncertainties on all derived quantities, such as the BH mass, are propagated through by assuming that the uncertainties are uncorrelated and independent. 

Because of its sensitivity to the flux in the wings of the line profile, care must be taken to define an appropriate range over which to measure the line dispersion. 
This is particularly true of Lorentzian-like profiles with extended wings.
In spectra of only moderate S/N the line limits are difficult to determine unambiguously, which introduces an extra degree of uncertainty in line dispersion measurements.  
In common with previous work \citep[e.g.][]{vestergaard06}, by default, the dispersion was calculated within $\pm 10\,000$\,\kms\, of the line centre, but this was extended when appropriate to avoid excluding a significant amount of line flux. 

\section{Results}
\label{sec:results}

A fundamental assumption on which single-epoch virial BH-mass estimates are based is that the widths of the broad emission lines are directly related to the virial motions of the emitting clouds moving in the gravitational potential of the central BH. 
However, the \civ line profiles of the quasars in our sample with the largest \civ blueshifts indicate that non-virial motions, very likely due to outflows, are having a significant effect on the observed \civ emission velocity profile \citep[e.g.][]{gaskell82,baskin05,sulentic07,richards11,wang13}.  
As shown in Fig.~\ref{fig:civ_composites}, at fixed emission-line EW, virtually the entire \civns-profile appears to shift blueward and the change in line shape is not simply an enhancement of flux in the blue wing of a still identifiable symmetric component. 
While gravity almost certainly plays a key role, determining the escape velocity for out-flowing material for example, it is clear that the virial assumption, on which single-epoch BH-mass measurements are predicated, is not straightforwardly applicable for the \civns-emission line in quasars exhibiting large blueshifts. 

The main aim of this paper is to investigate potential systematic trends in \civns-based single-epoch virial BH masses as a function of the \civ blueshift. 
Calibrations using \hb (and therefore also \hans) are generally accepted to be the most reliable, since most reverberation mapping employs the \hb line and the $R-L$ relation has been established using \hbns.
Therefore, we will test the reliability of the \civns-based estimates by comparing \civ line profiles to \ha profiles in the same quasars. 

\subsection{Characterising the emission-line profiles}
\label{sub:charemprof}

\begin{figure*}
	\includegraphics[width=2\columnwidth]{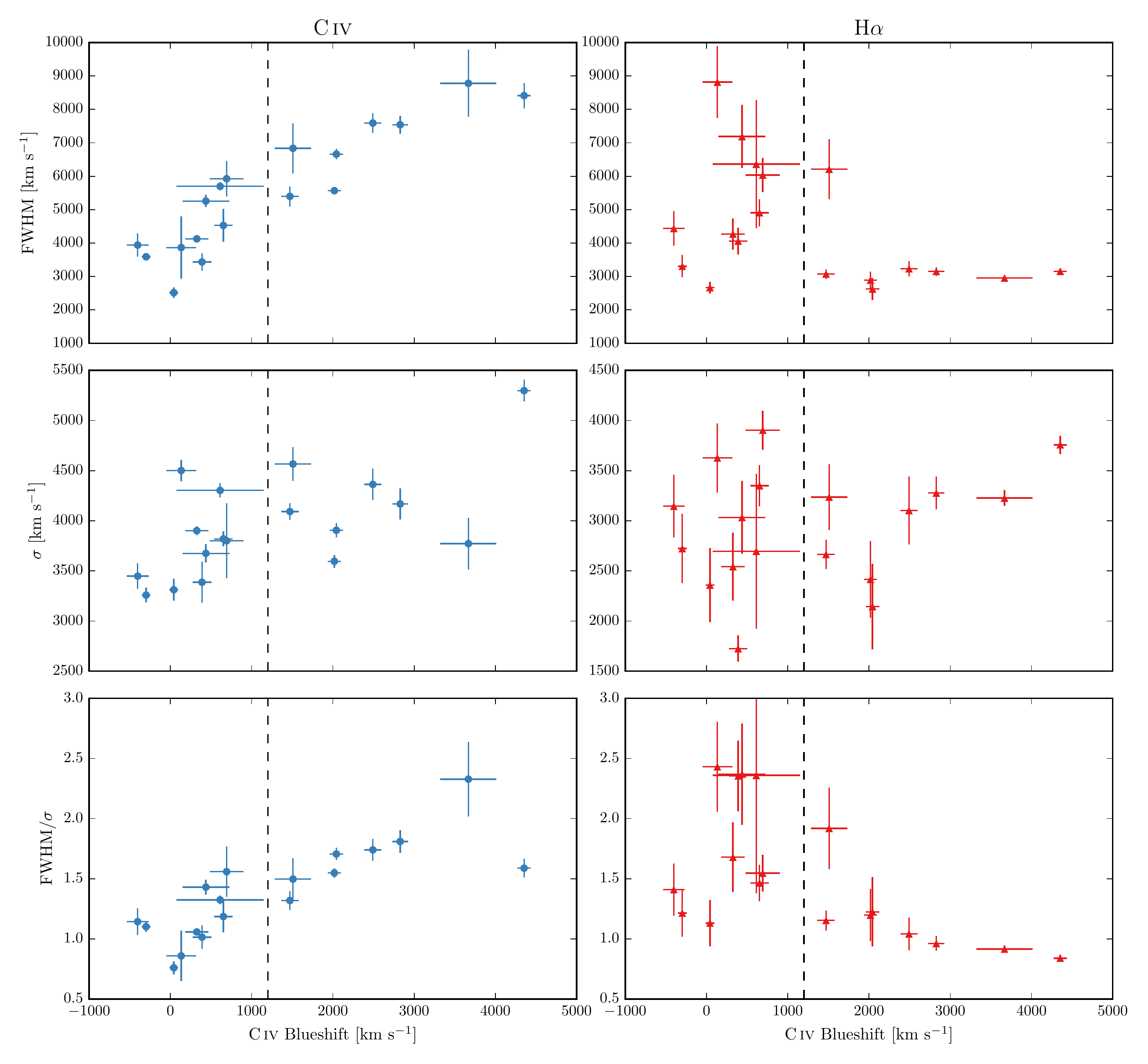} 
    \caption{The FWHM, dispersion ($\sigma$) and shape (FWHM/$\sigma$) of \civ and \ha as a function of the \civ blueshift. The vertical line demarcates the `high' and `low' \civ blueshift regimes discussed in the text. At high blueshift it is clear that BH masses estimated from the \civ FWHM (as is typically done at the redshifts considered) will be significantly larger than those estimated from the \ha FWHM.}
    \label{fig:line_comparison}
\end{figure*} 

There has been a considerable degree of attention paid to the effectiveness of different velocity-width measures of the \civns-emission; specifically, the line FWHM and the dispersion, $\sigma$, derived from the second-moment velocity \citep[e.g.][]{assef11, denney13}.
The FWHM and line dispersion trace different parts of the broad line velocity field, with the FWHM relatively more sensitive to any low-velocity core present and the line dispersion relatively more sensitive to the high velocity wings. 
The shape of the line can be characterised by the ratio FWHM/$\sigma$. 
FWHM/$\sigma$ $\simeq 2.35$ for a Gaussian profile, while FWHM/$\sigma$ $\simeq 1$ for a peakier Lorentzian profile\footnote{Strictly FWHM/$\sigma$ $\rightarrow 0$ for a Lorentzian profile, but values close to unity are typical when the dispersion is calculated over a velocity range, $\simeq\pm10\,000$\kms, used to parametrize broad emission lines in quasar spectra.}.
In practice, the line dispersion is almost certainly a more robust velocity indicator when the assumptions underlying the virial-origin of the emission-line velocity width are true and the spectral S/N and resolution are adequate.
This was demonstrated by \citet{denney13} for a sample of quasars possessing a significantly smaller range in \civns-blueshift than investigated here.

In reality, however, as highlighted by \citet{denney12}, contributions to the \civns-emission line profile from gas where virial motions do not dominate can be significant. 
Looking to the future, the results of the new reverberation-mapping projects \citep{shen15, kingoz15} will show what fraction of the \civns-emission line, as a function of velocity, does reverberate for quasars with an extended range of \civns-emission shapes. 
The derivation of quantitative corrections to transform velocity-width measures from single-epoch to reverberation-only line profiles should then be possible. 

As such information is not yet available, there is a strong rationale for investigating whether the systematic changes in the \civns-emission line profile can be used to improve the single-epoch BH-mass estimates derived using the \civ line. 
In the left panels of Fig.~\ref{fig:line_comparison} we show how the \civ FWHM, line dispersion, $\sigma$, and line shape, FWHM/$\sigma$, vary as a function of the blueshift. 
The \civ FWHM is correlated with the blueshift, with the median FWHM of quasars with the largest blueshifts a factor of 2-3 higher than quasars with only moderate blueshifts.
The dispersion, however, does not show a similarly strong systematic variation. 

Without knowledge of the \civns-blueshifts, the dynamic range present in the FWHM and line dispersion measurements accords with the expectations from the study of \citet{denney13}; the factor of $\simeq$4 spread in the FWHM measurements indicating greater sensitivity to the emission-line profile shape than is the case for the dispersion, which varies by a factor of only $\simeq$2. 
Adopting a value of 1200\,\kms\, to define `low' and `high' blueshift, the median \civns-emission dispersion for the low and high-blueshift samples differ by only 10 per cent. 
It follows, therefore, that while the dispersion provides a relatively line-profile independent measure of the velocity width for quasars where the underlying assumption regarding the virial-origin of the velocity width applies, quasars where the assumption is not true can be assigned apparently normal velocity-widths and hence potentially incorrect BH-masses. 

\begin{figure}
	\includegraphics[width=\columnwidth]{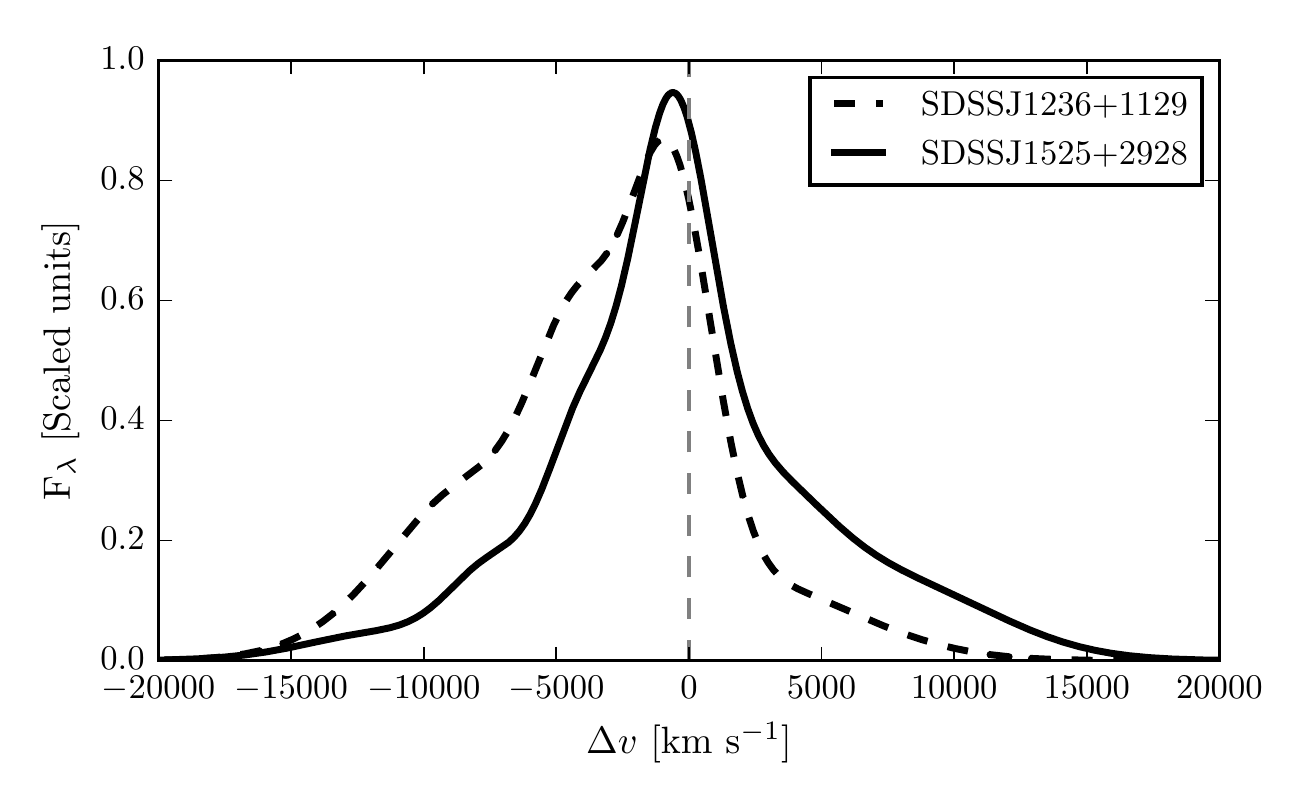} 
    \caption{Comparison of the \civ line profiles of SDSSJ1236+1129 and SDSSJ1525+0426. Notwithstanding the essentially identical dispersion values, the emission-line velocity fields differ dramatically and, therefore, the dispersion values cannot be measuring accurately the virial-induced velocity spread of the \civ emission in both quasars. }
    \label{fig:civ_comparison}
\end{figure}

To emphasise this point, in Fig.~\ref{fig:civ_comparison} we overlay the \civ line profiles of SDSSJ1236+1129 and SDSSJ1525+2928, whose dispersions (Table~\ref{tab:emissionproperties}) are indistinguishable (4168$\pm$271 and 4303$\pm128$\,\kms respectively). 
Notwithstanding the very similar dispersion values, the emission-line velocity fields differ dramatically and, therefore, the dispersion values cannot be measuring accurately the virial-induced velocity spread of the \civ emission in both quasars.

The analysis here, building on earlier work \citep[including][]{shen12, sulentic07}, confirms a link between \civ emission-line shape and blueshift, raising the prospect of developing a blueshift-dependent correction to single-epoch BH-mass estimates based on the \civ line. 
Expressed in another way, we are interested in testing if the significant systematic change in line shape as a function of \civ blueshift can be used to provide improved single-epoch BH-masses from the \civ emission line.  
The tightness of the correlation we observe between the \civ FWHM and blueshift implies that such an approach may be more effective than using the \civ emission-line velocity dispersion without reference to blueshifts.
A further practical advantage is that, given the typical S/N of current survey-quality spectra, virial BH mass estimates for high-redshift quasars are usually based on the FWHM rather than the dispersion \citep[e.g.][]{shen11}, which, being strongly affected by the continuum placement, is often found to be difficult to measure robustly \citep[e.g.][]{mejia-restrepo16}. 
As a first step towards the goal, below (Sec.~\ref{sec:hatrends}) we investigate the apparent systematic trends in the \ha FWHM and line shape as a function of \civ blueshift (shown in the right of Fig.~\ref{fig:line_comparison}).

\subsection{Computing BH mass estimates}
\label{sec:nonvirialmass}

Single-epoch virial BH mass estimates normally take the form

\begin{equation}
  \label{eq:virialmass}
  {\rm M_{BH}} = 10^{a} \left( \frac{\Delta V}{1000~{\rm km~s^{-1}}} \right)^b \left[ \frac{L_{\lambda}}{10^{44}~{\rm erg~s^{-1}}} \right]^c
\end{equation}

\noindent where $\Delta V$ is a measure of the line width (from either the FWHM or dispersion), $L_\lambda$ is the monochromatic continuum luminosity at wavelength $\lambda$, and $a$, $b$, and $c$ are coefficients, determined via calibration against a sample of AGN with reverberation-mapping BH mass estimates. Several calibrations have been derived using different lines (e.g. \hbns, \mgiins, \civns) and different measures of the line width (FWHM or dispersion) \citep[e.g.][]{vestergaard02,mclure02,vestergaard06,mcgill08,wang09,rafiee11,park13}.

Reverberation mapping measurements of nearby AGN have revealed the BLR to be stratified, with high-ionisation lines, including \civns, emitted closer to the BH than low-ionisation lines, including \ha and \hb \citep[e.g.][]{onken02}.
\citet{vestergaard06} found that the \civns-emitting region is at approximately half the radius of the \hbns/\ha emitting region.
Given the $\Delta V \propto R_{\rm BLR}^{-0.5}$ virial relation, this leads to the prediction that the \civ line widths should be $\simeq 1.4$ times broader than \ha for a given BH mass. 
More recently, \citet{denney12} found that there is a significant contribution from gas at larger radii to the \civ emission line, enhancing the profile at lower-velocity and leading to smaller FWHM or dispersion values. 
The ratio of the line widths is therefore predicted to be lower than the factor of $\simeq 1.4$. 

An alternate virial BH-mass calibration is proposed by \citet{park13}, using an improved sample of AGN with reverberation mapped masses. 
A major difference from the calibration of \citet{vestergaard06} is that \citet{park13}, recognising the poor correlation sometimes observed between the \civ and \hb FHWM, allow the exponent on the velocity width ($b$ in Eq.~\ref{eq:virialmass}) to vary.
Calibrating Eq.~\ref{eq:virialmass} against reverberation BH masses, they find a best-fit value of $b=0.56$, which is much less than the $b=2.0$ in the strict virial regime. 
As a result, the derived BH masses are much less sensitive to the \civns-emission line properties.
By contrast our approach is to investigate whether a more complete parametrization of the \civns-emission profile can be used to improve BH-mass estimates based on the conventional virial relation, with $b=2.0$.

\subsection{\civns-derived BH masses at low \civ blueshift}
\label{sec:bhm_lowbs}

The \ha and \civ FWHM (dispersion) of the 10 quasars with \civ blueshifts $<$1200\,\kms\, are linearly correlated, as expected if the dynamics of the BLR clouds are dominated by virial motions. 
The median \civns/\ha FWHM (dispersion) ratio is 0.91 (1.22) with standard deviation 0.17 (0.28). 
Thus, the dispersion-based comparison results in a median \civns/\ha consistent with the value of $\simeq1.4$ from assuming a virial origin for the emission but with a relatively large standard deviation. 
As predicted in Section~\ref{sec:nonvirialmass}, the FWHM-based comparison results in a systematically lower median \civns/\hans.
However, the correlation between the \civ and \ha FWHMs is significantly tighter, lending support to the proposal that corrections to BH-mass estimates based on the \civ emission line properties may be possible.

Virial BH masses were calculated using the widely adopted \citet{vestergaard06} calibrations. 
The \citet{vestergaard06} \civ FWHM calibration uses the monochromatic continuum luminosity at 1350\,\AA\, to predict the BLR radius and corresponds to ($a=6.66$, $b=2$, $c=0.53$) in Eq.~\ref{eq:virialmass}. 
The calibration coefficient $a=6.73$ in their equivalent dispersion-based relation. 
For the \hb calibration, \citet{vestergaard06} use the monochromatic continuum luminosity at 5100\,\AA\, and calibration coefficients corresponding to ($a=6.91$, $b=2$, $c=0.5$).
BH masses are computed using the line and continuum properties given in Tables \ref{tab:emissionproperties} and \ref{tab:continuumproperties}, and we convert our \ha emission-line velocity-width measures to predicted \hb widths using Eq.~\ref{eq:hb2hawidth}.

\begin{figure}
	\includegraphics[width=\columnwidth]{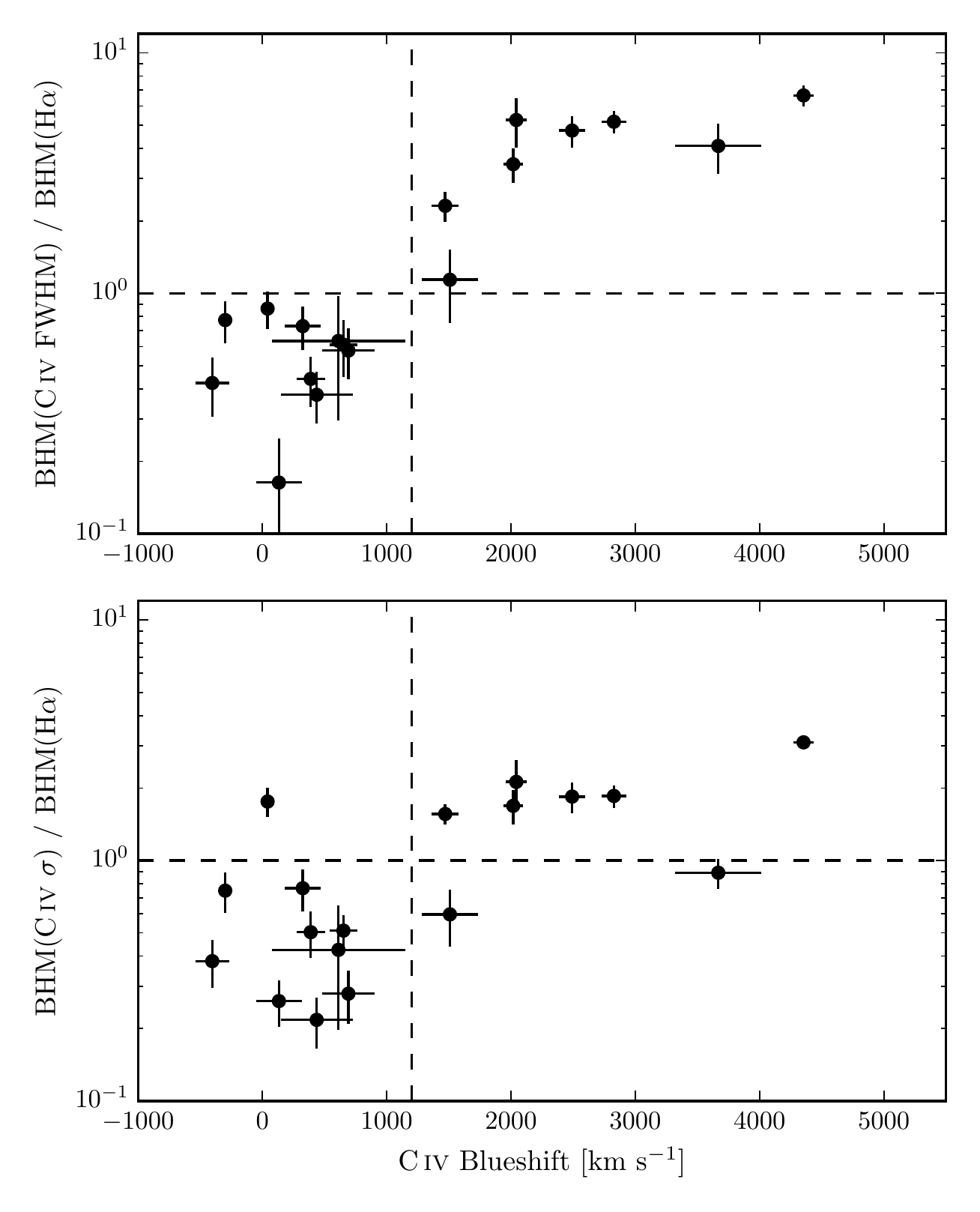}
    \caption{Comparison of virial BH mass estimates based on the \civ FWHM ({\it top}) and dispersion $\sigma$ ({\it bottom}) and \ha FWHM as a function of the \civ blueshift. The horizontal line indicates agreement between \civ and \ha BH masses, and the vertical line demarcates the `high' and `low' \civ blueshift regimes discussed in the text. The BH masses of quasars with moderate \civ blueshifts are underestimated when using the \civ FWHM, while the masses of quasars with large blueshifts are severely overestimated. This situation cannot be corrected by changing the exponent on the FWHM \citep[e.g.][]{rafiee11,park13} or the overall scaling in standard virial BH mass relations.}
    \label{fig:bhm_blueshift}
\end{figure}

As a direct consequence of the empirically small \civns/\ha FWHM ratio, the \civns-derived BH mass estimates are systematically lower than the corresponding \hans-derived masses when the blueshift is small.
This can be seen in Fig~\ref{fig:bhm_blueshift}, where for every quasar with a \civ blueshift $<$1200\,\kms (i.e. to the left of the dashed line), the \civns-derived BH mass is smaller than the corresponding \hans-derived mass.
The median fractional difference between the two estimates is 0.60.  

For the 10 quasars with low \civ blueshifts, we looked for correlations of the \civns/\ha FWHM ratio with other spectral properties.
We found marginal evidence for an anti-correlation with the \ha FWHM (Spearman coefficient 0.58 with p-value 0.08). 
Among the quasars with \ha FWHM $>4000$\kms we found the mean \civns/\ha FWHM ratio to be 0.83, compared to 1.01 for the quasars with \ha FWHM $<4000$\kms.  
Similar trends have been observed at low-$z$; in a sample of \citet{boroson92} quasars, \citet{baskin05} found the \civ line to be broader than \hb when the \hb FWHM $<$4000 \kms and narrower when the \hb FWHM $>$4000 \kms. 

\subsection{\civns-derived BH masses at high \civ blueshift}
\label{sec:highzmasses}

In Section~\ref{sub:charemprof} we have shown that the \civ emission at large \civns-blueshift is not dominated by virial-induced motions due to the BH. 
The empirically derived increase in the \civ emission FWHM with blueshift leads directly to an overestimate of BH-mass if the trend with blueshift is not taken into account. 
The availability of the \hans-spectra for the sample allows the quantification of the bias in inferred BH-mass under the assumption that the \ha emission line provides a reliable BH mass. 

Figure~\ref{fig:bhm_blueshift} shows the ratio of the \civns- and \hans-FWHM derived BH masses as a function of the \civ blueshift.  
We see that for quasars with \civ blueshifts $>2000$\,\kms, the \civns-based masses overestimate the \hans-based masses by as much as a factor of $\sim$5. 

The existence of a trend in the \civns-dispersion values with \civ blueshift is evident from inspection of the bottom left panel of Fig.~\ref{fig:line_comparison} but the systematic trend relative to the spread at fixed blueshift is significantly smaller than when using \civ FWHM. 
Similarly, Fig.~\ref{fig:bhm_blueshift} shows, at most, only a weak increase in the ratio of \civns- and \hans-derived masses. 
Without knowledge of the \civ blueshifts the distribution of \civns- and \hans-dispersion based BH masses could be taken to be reassuring. 
Including the \civns-blueshift information, however, demonstrates that any such interpretation is inherently flawed because the origin of the \civ emission velocity width is not due to virial-motions for a significant range of \civ blueshift. 
To reiterate the point made above (Sec.~\ref{sub:charemprof}), we believe that using a greater knowledge of the line profile (i.e. both the FWHM and blueshift) is a better motivated (and more practical) approach to obtaining more reliable virial BH mass estimates from the \civ line. 

The number of objects in our sample is small but an important factor contributing to the significant correlation evident in the FWHM version of Fig.~\ref{fig:bhm_blueshift} is a change in the emission-line shape of \ha as the \civns-blueshift increases.
By comparing the distributions of the \ha FWHM and dispersion as a function of \civns-blueshift (shown in the right-panels of Fig.~\ref{fig:line_comparison}), there is trend for the \ha lines to become peakier (with FWHM/$\sigma$ approaching unity) as the \civ blueshift increases. 
Whether the size of the true systematic bias in BH masses inferred from \civns-emission FWHM is as large as shown in Fig.~\ref{fig:bhm_blueshift} will depend on the future parametrization of the reverberation-component present in \hb (and \hans) profiles for quasars with high luminosities and large \civ blueshifts.

In summary, Fig.~\ref{fig:bhm_blueshift} illustrates the extent to which key derived physical parameters, including the BH mass and $L/L_{\rm Edd}$, could be systematically in error when \civns-FWHM measures are used without incorporating the information from the \civ blueshifts. 
Other authors have proposed empirical corrections to \civns-based BH masses based on similar systematic trends seen in the \civ line shape \citep{denney12} and the continuum-subtracted peak flux ratio of the ultraviolet emission-line blend of \siivns+\oiv (at 1400\,\AA) to that of \civ \citep{runnoe13}.
In Section~\ref{sec:discussionbiases} we apply these corrections the quasars in our sample, and discuss the effect they have on the systematic bias seen in Fig.~\ref{fig:bhm_blueshift}. 

\subsection{Population trends with \civ blueshift}
\label{sec:hatrends}

\begin{figure}
	\includegraphics[width=\columnwidth]{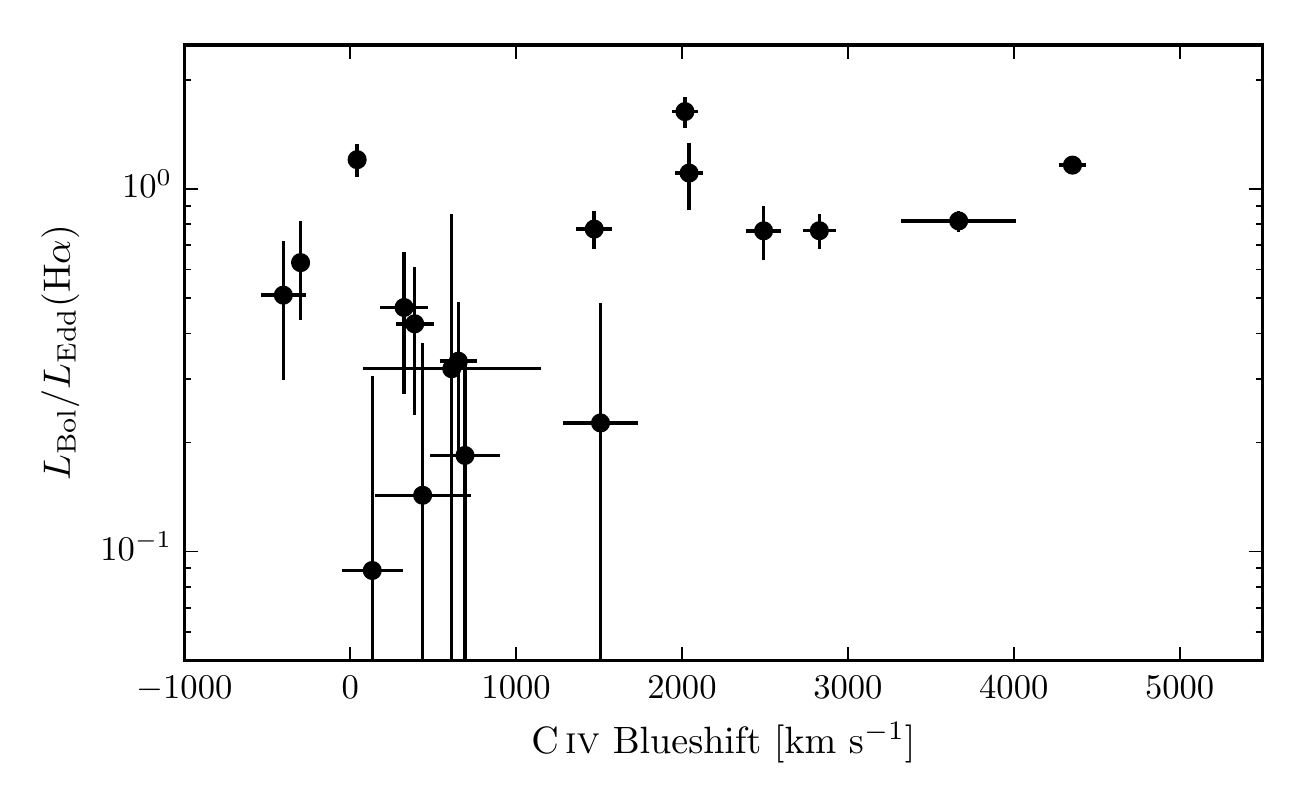}
    \caption{\hans-derived Eddington ratio versus \civ blueshift. At blueshift $\gtrsim$ 2000\kms\, all quasars have high accretion rates ($L/L_{\rm Edd} \simeq 1$). This is in agreement with \citet{kratzer15}, but in contrast to what one would derive from naive use of \civns-based BH mass scaling relations.}
    \label{fig:ha_edd_civ_bs}
\end{figure}

Even with the caveat regarding the small sample size, the differences in the \ha emission-profile as a function of \civns-blueshift (Fig.~\ref{fig:gridspectra_1}) appear to be systematic.
At \civns-blueshift $<$1200\,\kms, the \ha FWHM range is $\simeq2700 - 8800$\,\kms, with mean $\simeq$5200\,\kms.
However, amongst the six quasars with \civns-blueshift $>$2000\,\kms, the mean \ha FWHM=3000\,\kms, with a scatter of just 200\,\kms. 
The apparent trend of peakier \hans-emission, with FWHM/$\sigma$ close to unity, at large \civns-blueshift is enhanced by the modest increase in \ha EW with blueshift (Table~\ref{tab:emissionproperties}). 
Amongst the low-\civns-blueshift population there are in addition quasars with broader and more Gaussian-like \ha line profiles, with FWHM/$\sigma \simeq 2$ . 

The change in the \ha emission-line profiles as a function of \civns-blueshift means that the \hans-FWHM derived BH masses at high-blueshift are smaller than the sample mean. 
We transformed the observed luminosity into a mass-normalised accretion rate (Eddington ratio).
To convert the monochromatic luminosity, which is observed, in to a bolometric luminosity we use the bolometric correction factor given by \citet{richards06} ($L_{\rm bol} = 9.26L_{5100}$).
Although there is evidence that the bolometric correction factor is a function of the luminosity, as well as of other parameters including the \civ blueshift \citep{krawczyk13}, the differences are small over the parameter range covered by our sample, and for simplicity we adopt a constant factor. 

The results, shown in Fig.~\ref{fig:ha_edd_civ_bs}, show that at large blueshifts quasars are accreting at around their Eddington limits (Fig.~\ref{fig:ha_edd_civ_bs}). 
The converse is, however, not true, i.e. not all quasars with high Eddington ratios possess large \civ blueshifts \citep[see][]{baskin05}.

\section{Discussion}
\label{sec:discussion}

\subsection{Biases in single-epoch \civns-based BH-mass estimates}
\label{sec:discussionbiases}

The \civ line profiles of the quasars with  the largest \civ blueshifts (in the bottom right of Fig.~\ref{fig:civ_space}) demonstrate that non-virial motions are having a significant effect on the \civ BLR dynamics. 
At fixed emission-line EW, almost the entire \civns-profile appears to shift blueward and the change in line shape is not simply an enhancement of flux in the blue wing of a symmetric component. 
It is clear that the virial assumption, on which single-epoch BH-mass measurements are predicated, is not straightforwardly applicable for the \civns-emission line in quasars exhibiting large blueshifts. 

Quantitatively, the \civns-shape change is most apparent from the FWHM values, which are strongly correlated with the \civns-blueshift. 
This trend has previously been identified, by comparison with \mgii at lower-redshifts \citep{shen08,shen11} and \hb at higher redshifts \citep{shen12}.
We find that virial BH mass estimates based on the \civ FWHM will overestimate the true mass by a factor of $\sim$5 for objects exhibiting the largest \civ blueshifts. 

\begin{figure}
	\includegraphics[width=\columnwidth]{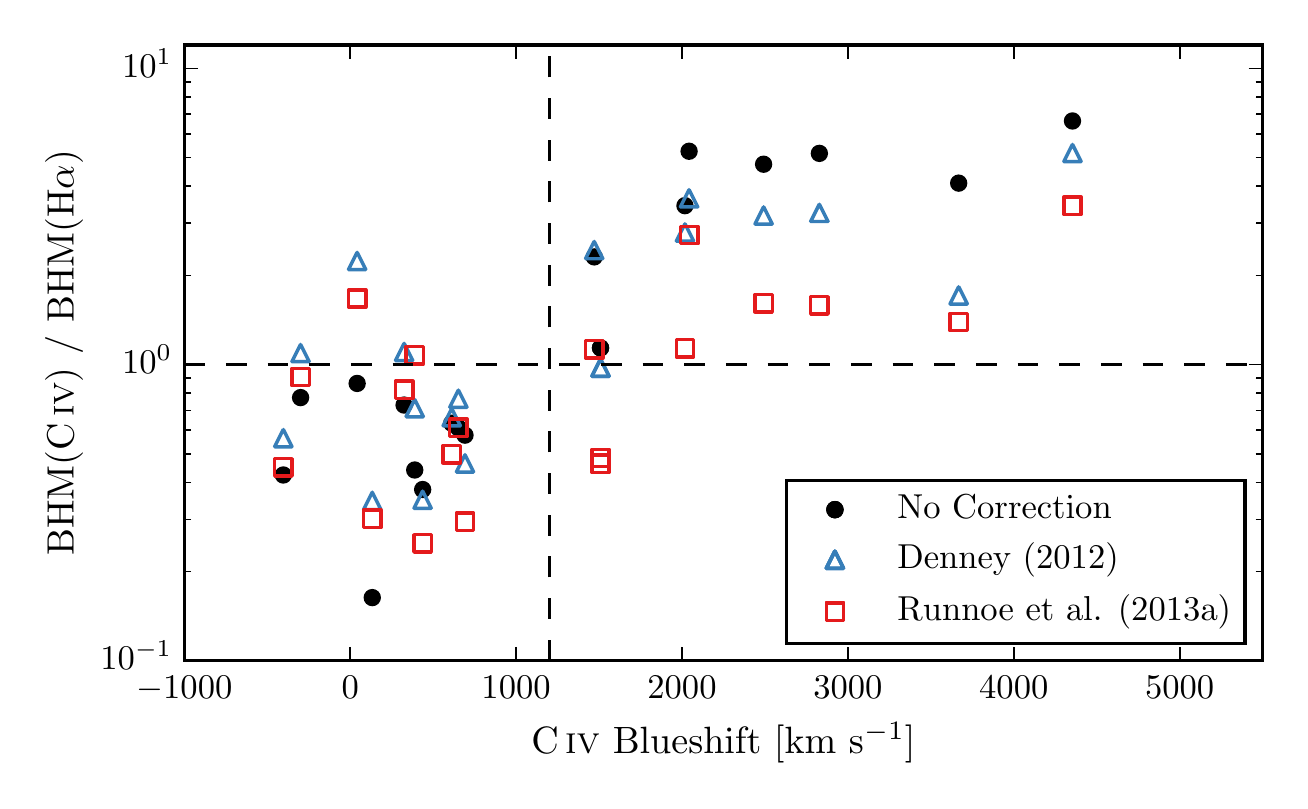}
    \caption{Comparison of BH mass estimates derived from the FWHM of \civ and \ha as a function of the \civ blueshift (\textit{black circles}), and after applying corrections to the \civns-derived mass based on the line shape \citep[\textit{blue triangles};][]{denney12} and the peak flux ratio of the \siivns+\oiv blend relative to \civ \citep[\textit{red squares};][]{runnoe13}. While the shape-based correction improves the consistency between BH mass estimates in the low-blueshift population, only the \siivns+\oivns/\civ peak flux-based correction is effective at high blueshifts (although a weak systematic remains).}
    \label{fig:corrections}
\end{figure}

In contrast, the \civ line dispersion does not show a similarly strong dependence on the blueshift. 
This is a result of the shape of the \civ line profile being dependent on the blueshift; the low-blueshift profiles are peakier (FWHM/$\sigma \simeq 1$) than the high-blueshift profiles (FWHM/$\sigma \simeq 2$). 
\citet{denney12} found the level of contamination in single-epoch spectra from non-reverberating gas to be correlated with the shape (FWHM/$\sigma$) of the \civ profile. 
Their investigation was based on a sample of reverberation mapped quasars, which have a narrow range of \civns-emission line shapes, including the absence of any objects with large \civ blueshifts. 
The FWHM/$\sigma$-based correction to the \civ FWHM proposed by \citet{denney12} increases the inferred BH masses by $\sim$0.2 dex at the low end of our \civ blueshift distribution, thereby reducing part of the systematic trend in the BH mass (Fig.~\ref{fig:corrections}).  
However, it is not applicable at the high \civ blueshift end, where velocity widths are likely dominated by non-virial motions.
Based on the typical line shape of \civ in these high-blueshift quasars (FWHM/$\sigma \simeq 2$), the \citet{denney12} correction decreases the predicted BH masses by $\sim$0.3 dex, which has only a moderate decreasing effect on the strong systematic (Fig.~\ref{fig:corrections}).  

While the \civns-line dispersion is largely independent of the blueshift, it does not follow that dispersion-based BH-mass estimates are correct, because the underlying assumption regarding the virial-origin of the \civ emission profile breaks down at large blueshifts.
Furthermore, given the difficulty in obtaining reliable dispersion measurements from survey-quality spectra with limited S/N \citep[e.g.][]{mejia-restrepo16}, virial BH-mass estimates for existing large samples of high-redshift quasars are usually based on the FWHM \citep[e.g.][]{shen11}. 
Our work therefore suggests that a viable recipe for obtaining more reliable BH mass estimates for large numbers of quasars at high redshift is to measure both the FWHM and the blueshift, which together can be used to derive a FWHM corrected for the non-virial contribution. 

Although we do not have enough quasars in our sample to derive a reliable quantitative correction to BH-masses as a function of \civ blueshift, we are assembling a large sample of quasar spectra with coverage of both \civ and \hbns/\ha in order to empirically validate such a correction. 
Both \citet{runnoe13} and \citet{shen12} considered a similar approach, but concluded that a low-ionisation broad line (e.g. \mgiins), or features from the quasar NLR, are required to determine the systemic redshift and hence the \civ blueshift. 
Empirical tests of the reliability of the improved \citet{hewett10} redshifts for the SDSS DR7 quasars \citep{shen16} and the availability of the largely SED-independent principal component analysis redshifts for DR10 and DR12 \citep{paris14, paris16} already allow meaningful corrections to BH-mass estimates for quasars exhibiting large \civns-blueshifts.
Our intention is, however, to determine a definitive correction formula using the redshifts from Allen \& Hewett (2016, in preparation) for both DR7 and DR12.

Given the difficulty of measuring reliable \civ blueshifts, \citet{runnoe13} opted instead to use the continuum-subtracted peak flux ratio of the ultraviolet emission-line blend of \siivns+\oiv (at 1400\,\AA) to that of \civ to correct for non-virial contributions to the \civ velocity width. 
This parameter was chosen because it showed the strongest correlation with the FWHM \civns/\hb residuals, as well as with the strengths of optical \oiii and \feiins. 
The strengths of optical \oiii and \feiins, being parameters in EV1, are also correlated with the \civ blueshift \citep{sulentic07}. 
As the \civ blueshift increases the EW decreases systematically (Fig.~\ref{fig:civ_space}).
The \siivns+\oiv emission-line blend, however, shows significantly less systematic variation. 
Therefore, while the \citet{runnoe13} \siivns+\oivns-based correction is effective in practice, the \civ blueshift measurement provides a more direct measure of the non-virial contributions to the \civ velocity width.

In our sample, we find the 1400\,\AA/\civ peak flux ratio to be strongly correlated to the \civ blueshift (the Spearman coefficient for the correlation is 0.82, $p$-value 2{\sc e}-5).
As such, the correction to \civns-based virial masses proposed by \citet[][their equation 3]{runnoe13} removes a large part of the systematic in the \hans/\civ FWHM residuals with the \civ blueshift (Fig.~\ref{fig:corrections}); the median \civns/\ha FWHM ratio at large \civ blueshifts ($>$1200\kms) is reduced from 4.5 to 1.5.
However, at low ($<$1200\kms) \civ blueshifts, the trend for \civ to predict lower BH masses persists, and the scatter between the \civns- and \hans-based masses increases by a factor of two.   
In accordance with our expectations, we find the FWHM \civns/\ha residuals to be more tightly correlated to the \civ blueshift (Spearman coefficient 0.82, $p$-value 3{\sc e}-5) than to the 1400\,\AA/\civ peak flux ratio (Spearman coefficient 0.72, $p$-value 7{\sc e}-4). 

\subsection{Possible systematic trends in \ha BH-mass estimates}

In Section~\ref{sec:hatrends}, we found that the quasars with large \civ blueshifts have systematically narrower \ha FWHM (Fig.~\ref{fig:line_comparison}).
Using the standard virial BH mass calibrations this implies that the high \civ blueshift population have high accretion rates ($L/L_{\rm Edd} \simeq 1$; Fig.~\ref{fig:ha_edd_civ_bs}). 
This interpretation requires some caution since the emission-line shape (characterized by the value of FWHM/$\sigma$) of \ha is also changing as a function of the \civ blueshift (Fig.~\ref{fig:line_comparison}). 
At low \civ blueshifts there are a range of shapes, but all of the quasars exhibiting large \civ blueshifts have peaky \ha profiles with FWHM/$\sigma \simeq 1$. 
This raises the question of whether the \ha FWHM is a reliable proxy for the virial-induced velocity dispersion for the full range of \ha line shapes we have in our sample. 

When calibrating the virial-product to masses derived independently using the BH mass \---\ stellar velocity dispersion ($M_\odot-\sigma$) relation, \citet{collin06} find that the scaling factor, $f$, is a factor $\sim2$ larger for their Population `1' sources (with FWHM/$\sigma < 2.35$ and essentially equivalent to population A of Sulentic and co-workers and to the high-blueshift quasars here) than for their Population 2 (with FWHM/$\sigma > 2.35$). 
For single-epoch BH-mass estimates, assuming a constant value of $f$, as is normally done \citep[e.g.][]{vestergaard06}, means that Population 1 masses will be underestimated and Population 2 will be overestimated.
In the context of this result from \citet{collin06}, our high-blueshift objects all possess peaky \hans-lines and, while our quasar sample probes much higher luminosities and masses, the true BH-masses may also be underestimated.
Adopting such an interpretation, the amplitude of the trend seen in Fig.~\ref{fig:bhm_blueshift} might not be so pronounced.

As mentioned in Section~\ref{sec:introduction} and discussed in \citet{richards11}, quasars with current reverberation mapping measurements have a restricted range of \civns-line shapes. 
There are currently very few reverberation-mapping measurements of quasars with large \civ blueshifts but the results of the large on-going statistical reverberation mapping projects \citep[e.g.][]{shen15} for luminous quasars at high-redshift will go some way to establishing whether the quasar broad line regions producing Balmer emission look the same for objects with very different \civns-emission blueshifts. 

Although the EV1-trends \citep{sulentic00b,shen14} are most likely driven by the accretion rate, orientation may also have a role to play in determining the observed properties of the BLR. 
\citet{shen14} argue that a large part of the scatter observed in the \hb FWHM relates not to a spread in BH masses, but rather to the orientation of the BLR relative to the line-of-sight.
For this to be true, the BLR would need to be in a flattened disc-like geometry, in which case the observed line width would increase with the inclination of the disc relative to the line of sight. 
\citet{brotherton15b} found that the core-dominance of radio-loud quasars, which is believed to be a reliable proxy for orientation, at least in a statistical sense, is significantly correlated with the \hb FWHM and hence with the BH-mass estimates. 
This raises the question of whether the narrow \ha emission lines observed in the quasars with the largest \civ blueshifts could be an orientation effect. 
However, there is no evidence that the \civ blueshift is dependent on the orientation \citep[inferred from the radio core-dominance;][]{richards11,runnoe14}. 
Furthermore, \citet{leighly04} showed that the \heiins\l1640 emission-line properties of quasars with large \civ blueshifts are more consistent with differences in the SED rather than differences in the orientation.
\citet{collin06} showed that orientation effects were also sub-dominant to the Eddington ratio in determining the shape of the \hb line and
the \ha line shape trend we observe is consistent with the finding of \citet{marziani03} that the \hb emission profiles of high/low Eddington ratio low-$z$ quasars and type 1 Seyfert nuclei are well fit by Lorentzian and double Gaussian profiles respectively.  
Overall, therefore, orientation does not appear to be the dominant effect in determining the \civ blueshift and correlated changes in the \ha line profile. 

\subsection{Accretion-rate trends in the quasar population}

The blueshifting of \civ is usually interpreted as evidence for strong outflows resulting from the presence of a radiation-driven accretion-disc wind. 
\citet{richards02} found that quasars with large \civ blueshifts have weak \heiins.
This is evidence for weak soft X-ray continuum emission \citep{leighly04}, which would allow a strong line-driven wind to form.  
The strength of such a wind is predicted to be related to the quasar far-ultraviolet SED, which, in turn, could be related to the mass-accretion rate.
This picture is therefore consistent with our finding that the quasars with large \civ blueshifts have high accretion rates. 

All of the objects in our sample which exhibit large \civ blueshifts would be classified as population A in the \citet{sulentic00} scheme based on the \ha FWHM (see Section~\ref{sec:introduction}). 
Our results therefore support the idea of the \citet{sulentic00} A/B division being driven by the Eddington ratio, with population A sources possessing higher accretion rates.
However, we also observe a number of quasars which have high Eddington ratios but do not have line profiles suggestive of strong outflows in the \civ BLR.  
This suggests that a high accretion rate is a necessary but not sufficient condition for the existence of outflows \citep{baskin05}. 

The two-dimensional nature of the \civ emission line parametrization and the apparent anti-correlation between \civ EW and \civ blueshift suggests that the quasar population exhibits a continuum of properties. 
As such, more accurate \civ blueshift measurements for SDSS-quasars should allow an improved mapping between the \civns-emission properties and key physical parameters of the quasars.
This includes improving our understanding of the origin of quasars with exceptionally weak, blueshifted \civ emission \citep[weak emission line quasars;][]{luo15} which could be exotic versions of wind-dominated quasars \citep{plotkin15}.

\subsection{The BAL parent population}

Classical high-ionization BAL (HiBAL) quasars are also predominantly Population A objects in the scheme of \citet{sulentic00}. 
There are no HiBAL quasars in our sample by design (Section~\ref{sec:selection}) but it is generally accepted that quasars which show high-ionisation BALs are likely to be radiating with relatively high $L/L_{\rm Edd}$ \citep[e.g.][]{zhang14}. 
We therefore propose that the subset of the quasar population that exhibits large \civns-emission blueshifts, with high-EW and narrow-\ha emission lines, may be directly related to the HiBAL quasar population \---\ perhaps even the `parent' population \citep{richards06conf}. 
A prediction of such a linkage is that near-infrared observations of the rest-frame optical spectra of HiBAL quasars will show strong, relatively narrow, Balmer emission lines, very similar to those of the quasars with high \civns-blueshifts presented in this paper \citep[see][for such a study]{runnoe13b}. 

\subsection{The frequency of quasars with high accretion rates}

Quantifying the frequency of quasars producing outflows as a function of key parameters, e.g. quasar luminosity, BH-mass, redshift,... will be important to constrain models of quasar-galaxy evolution.  
At fixed BH mass, the intrinsic and the observed fraction of quasars exhibiting properties that depend on the Eddington ratio can differ significantly. 
As an illustration, we consider the implications for the intrinsic fraction of quasars possessing large \civ blueshifts given the observed numbers in the $m_i<19.1$ flux-limited sub-sample of the SDSS DR7 quasar catalogue, from which our quasar sample is effectively drawn (Section~\ref{sec:selection}). 
In order to estimate the size of the selection effect, we considered the detection probability for a much-simplified quasar population. 
We assume that all quasars with \civ blueshifts $>$1200\,\kms have enhanced accretion rates relative to the `normal' population (with \civ blueshifts $<$1200\,\kms). 
If the accretion rate of the high-blueshift population is double the rate of the low-blueshift population (which is true in an average sense \---\ see Fig.~\ref{fig:ha_edd_civ_bs}), then the high-blueshift population will be brighter by $\simeq$0.75 magnitude.
Under the assumption that the BH mass distribution is independent of the \civ blueshift, the high-blueshift population will then be over-represented in a flux-limited sample.
To estimate the size of the bias, we need to know how many more quasars, at redshifts $2 < z < 2.5$, there are with $m_i<19.1+0.75=19.85$ relative to $m_i < 19.1$.
This is the fraction of the population which, as a consequence of having enhanced accretion rates, are boosted above the survey flux limit.    
The main colour-selected SDSS DR7 quasar catalogue extends only to $m_i= 19.1$ and, assuming the luminosity function is continuous\footnote{The luminosity function and number-counts vary only smoothly \citep[e.g.][]{ross13} for the magnitude and redshift range used here.} we thus use the number counts at $m_i < 19.1$ and $m_i < 18.35$, which differ by a factor of $\simeq 4$. 

At redshifts $2 < z <2.5$, there are 3,834 quasars with \civ blueshifts $<$1200\,\kms and 2,484 with blueshifts $>$1200\,\kms in the SDSS DR7 $m_i < 19.1$ quasar sample, a ratio of $\sim$2:1. 
The above calculation, although much idealised, suggests that the intrinsic fraction of high-blueshift quasars is a factor of four smaller than in the flux-limited sample (i.e. $\sim$15\,per cent of the ultraviolet-selected non-BAL quasar population). 

\section{Conclusions}
\label{sec:conclusions}

We have presented an analysis of biases in \civns-derived virial BH masses of high-luminosity (L$_{\rm{bol}} \sim$ 47 erg s$^{-1}$) quasars at redshifts $\sim2.5$ from the Sloan Digital Sky Survey. 
Many authors have reported a large scatter between \civns- and \hans/\hbns-based masses, and part of this scatter has been shown to correlate with the \civ blueshift \citep{shen12}.
The blueshifting of \civ is usually interpreted as evidence for strong outflows which, most likely, result from the presence of a radiation line-driven accretion-disc wind. 
Our study is the first to examine this bias systematically across the full range of \civns-emission line blueshifts observed in the SDSS sample. 
In particular, we have used rest-frame optical spectra of 19 quasars in the redshift range $2 < z < 2.7$ to directly compare \civ and \ha emission properties as a function of the \civ blueshift. 
We reach the following conclusions:

\begin{itemize}

\item{A strong correlation between \civns-velocity width and blueshift is found and at large blueshifts, $>$2000\,\kms, the velocity-widths are dominated by non-virial motions. 
This suggests that the assumption that velocity-widths arise from virial-induced motions, on which single-epoch BH-mass measurements are predicated, is not straightforwardly applicable to these high-blueshift quasars.}

\item{We use the \ha emission line to provide BH-mass estimates that are unbiased by non-virial contributions to the velocity width. 
We find that the \civns-based BH masses of quasars with low \civ blueshifts are systematically underestimated (by a factor of $\sim$1.7) whereas the masses of quasars with large blueshifts are severely overestimated (by a factor of $\sim$5).} 

\item{We find a systematic change in the shape of the \ha line profile as a function of the \civ blueshift. 
Specifically, the \ha line profiles of the quasars with high \civ blueshifts are all `peaky' with FWHM/$\sigma$ close to unity.} 

\item{We suggest that the high \civ blueshift quasars are high Eddington-ratio objects that are inherently rare (comprising $\sim$15\% of the UV-selected sample), but are being boosted in number by a factor of $\sim$4 in the flux-limited SDSS sample.}

\end{itemize}

With a relatively small sample of 19 quasars we have been able to uncover systematic trends in the \civ and \ha emission line shapes as a function of the \civ blueshift.
This confirms the prospect of developing a blueshift-dependent correction to \civns-based single-epoch BH-mass estimates using a larger samples of luminous quasars with both rest-frame UV and rest-frame optical spectroscopy. 
We are currently in the process of assembling such a sample, which will contain $\sim$300 luminous quasars, 80\, per cent at redshifts $z\geq$2.    
A new SED-independent redshift-estimation algorithm (Allen \& Hewett 2016, in preparation) makes it possible to quantify the distribution of \civns-blueshifts in the observed quasar population as a whole, thereby allowing us to make empirical corrections to \civns-based BH-masses for all luminous, high-redshift SDSS/BOSS quasars.

\section*{Acknowledgements}

We thank the anonymous referee for a careful and constructive report that resulted in significant improvements to the paper. 

LC thanks the Science and Technology Facilities Council (STFC) for the award of a studentship. MB and PCH acknowledge support from the STFC via a Consolidated Grant to the Institute of Astronomy, Cambridge. MB acknowledges support from STFC via an Ernest Rutherford Fellowship. 

The paper is based on observations made with the 4.2 m William Herschel Telescope operated on the island of La Palma by the ING at the Observatorio del Roque de los Muchachos of the Instituto de Astrof\'isica de Canarias.

Funding for the SDSS and SDSS-II has been provided by the Alfred P. Sloan Foundation, the Participating Institutions, the National Science Foundation, the U.S. Department of Energy, the National Aeronautics and Space Administration, the Japanese Monbukagakusho, the Max Planck Society, and the Higher Education Funding Council for England. The SDSS Web Site is http://www.sdss.org/.

The SDSS is managed by the Astrophysical Research Consortium for the Participating Institutions. The Participating Institutions are the American Museum of Natural History, Astrophysical Institute Potsdam, University of Basel, University of Cambridge, Case Western Reserve University, University of Chicago, Drexel University, Fermilab, the Institute for Advanced Study, the Japan Participation Group, Johns Hopkins University, the Joint Institute for Nuclear Astrophysics, the Kavli Institute for Particle Astrophysics and Cosmology, the Korean Scientist Group, the Chinese Academy of Sciences (LAMOST), Los Alamos National Laboratory, the Max-Planck-Institute for Astronomy (MPIA), the Max-Planck-Institute for Astrophysics (MPA), New Mexico State University, Ohio State University, University of Pittsburgh, University of Portsmouth, Princeton University, the United States Naval Observatory, and the University of Washington.



\bibliographystyle{mn2e}
\bibliography{bibliography}





\bsp	
\label{lastpage}
\end{document}